\documentclass[apj]{emulateapj}

\usepackage{epsfig}
\usepackage{amsmath}
\usepackage{natbib}
\usepackage{graphicx,subfigure}
\usepackage{calc}
\usepackage{verbatim}
\usepackage{tabularx}
\usepackage{xcolor}

\newcommand{\ztwo}{$z\sim2 \hspace{4pt}$}

\newcommand{\ls}{\hspace{2pt}}
\newcommand{\ha}{H$\alpha$\ls}				    
\newcommand{\hb}{H$\beta$\ls}					    
\newcommand{\nii}{[N{\small II}]\ls}                                       
\newcommand{\sii}{[S{\small II}]\ls}                                       

\newcommand{\oiii}{[O{\small III}]}
\newcommand{\niiha}{[N{\small II}]/\ha}				    
\newcommand{\siiha}{[S{\small II}]/\ha}				    

\newcommand{\oiiihb}{[O{\small III}]/\hb}

\newcommand{\msun}{M$_{\odot}$} 			   

\newcommand{\mstar}{M$_{*}$} 				    

\newcommand{\lam}{$\lambda$}

\newcommand{\sigsfr}{$\Sigma_{SFR}$}

\newcommand{\msunyr}{\msun\ yr$^{-1}$}		    
\newcommand{\kms}{km~s$^{-1}$}			              

\shorttitle{Nebular excitation in \ztwo SFGs}
\shortauthors{S. F. Newman et al.}

\begin{document}

\title{Nebular Excitation in \ztwo Star-forming Galaxies from the SINS and LUCI Surveys: \\ The Influence of Shocks and AGN$^{*}$}
\author{Sarah F. Newman\footnotemark[1,17], Peter Buschkamp\footnotemark[2], Reinhard Genzel\footnotemark[1,2,3], Natascha M. F\"orster Schreiber\footnotemark[2], Jaron Kurk\footnotemark[2],  Amiel Sternberg\footnotemark[4], Orly Gnat\footnotemark[5], David Rosario\footnotemark[2], Chiara Mancini\footnotemark[6],  Simon J. Lilly\footnotemark[7], Alvio Renzini\footnotemark[6], Andreas Burkert\footnotemark[8],  C. Marcella Carollo\footnotemark[7], Giovanni Cresci\footnotemark[9], Ric Davies\footnotemark[2], Frank Eisenhauer\footnotemark[2], Shy Genel\footnotemark[10],  Kristen Shapiro Griffin\footnotemark[11], Erin K. S. Hicks\footnotemark[12], Dieter Lutz\footnotemark[2], Thorsten Naab\footnotemark[13], Yingjie Peng\footnotemark[7],  Linda J. Tacconi\footnotemark[2], Stijn Wuyts\footnotemark[2],  Gianni Zamorani\footnotemark[14], Daniela Vergani\footnotemark[15] and Benjamin J. Weiner\footnotemark[16] \\ 
\scriptsize{
$^{1}$Department of Astronomy, Campbell Hall, University of California, Berkeley, CA 94720, USA \\
$^{2}$Max-Planck-Institut f\"ur extraterrestrische Physik (MPE), Giessenbachstr.1, D-85748 Garching, Germany \\
$^{3}$Department of Physics, Le Conte Hall, University of California, Berkeley, CA 94720, USA \\
$^{4}$School of Physics and Astronomy, Tel Aviv University, Tel Aviv 69978, Israel \\ 
$^{5}$Racah Institute of Physics, The Hebrew University, Jerusalem 91904, Israel \\
$^{6}$Osservatorio Astronomico di Padova, Vicolo dell'Osservatorio 5, Padova, I-35122, Italy \\
$^{7}$Institute of Astronomy, Department of Physics, Eidgen\"ossische Technische Hochschule, ETH Z\"urich, CH-8093, Switzerland \\
$^{8}$Universit\"ats-Sternwarte Ludwig-Maximilians-Universit\"at (USM), Scheinerstr. 1, M\"unchen, D-81679, Germany \\
$^{9}$Istituto Nazionale di AstrofisicaÐOsservatorio Astronomico di Arcetri, Largo Enrico Fermi 5, I Ð 50125 Firenze, Italy \\
$^{10}$Harvard-Smithsonian Center for Astrophysics, 60 Garden Street, Cambridge, MA 02138 USA \\
$^{11}$Space Sciences Research Group, Northrop Grumman Aerospace Systems, Redondo Beach, CA 90278, USA \\
$^{12}$Department of Astronomy, University of Washington, Box 351580, U.W., Seattle, WA 98195-1580, USA \\
$^{13}$Max-Planck Institute for Astrophysics, Karl Schwarzschildstrasse 1, D-85748 Garching, Germany \\
$^{14}$INAF Osservatorio Astronomico di Bologna, Via Ranzani 1, 40127 Bologna, Italy \\
$^{15}$INAF Istituto di Astrofisica Spaziale e Fisica Cosmica di Bologna, Via P. Gobetti 101, 40129 Bologna, Italy \\
$^{16}$Steward Observatory, University of Arizona, Tucson, AZ 85721, USA \\
$^{17}$email: sfnewman@berkeley.edu
}}

\footnotetext[*]{Based on observations at the Very Large Telescope (VLT) of the European Southern Observatory (ESO), Paranal, Chile (ESO program IDs 073.B-9018, 076.A-0527, 079.A-0341, 080.A-0330, 080.A-0339, 080.A-0635, 083.A-0781,084.A-0853, 087.A-0081, 091.A.-0126) and at the Large Binocular Telescope (LBT) on Mt. Graham in Arizona.}



\begin{abstract}
Based on high-resolution, spatially resolved data of 10 \ztwo star-forming galaxies from the SINS/zC-SINF survey and LUCI data for 12 additional galaxies, we probe the excitation properties of high-z galaxies and the impact of active galactic nuclei (AGN), shocks and photoionization. We explore how these spatially-resolved line ratios can inform our interpretation of integrated emission line ratios obtained at high redshift. Many of our galaxies fall in the `composite' region of the z $\sim$ 0 \niiha \ls versus \oiiihb \ls diagnostic (BPT) diagram, between star-forming galaxies and those with AGN. Based on our resolved measurements, we find that some of these galaxies likely host an AGN, while others appear to be affected by the presence of shocks possibly caused by an outflow or from enhanced ionization parameter as compared with HII regions in normal local star-forming galaxies. We find that the Mass-Excitation (MEx) diagnostic, which separates purely star-forming and AGN hosting local galaxies in the \oiiihb \ls versus stellar mass plane, does not properly separate \ztwo galaxies classified according to the BPT diagram. However, if we shift the galaxies based on the offset between the local and \ztwo mass-metallicity relation (i.e. to the mass they would have at z $\sim$ 0 with the same metallicity), we find better agreement between the MEx and BPT diagnostics. Finally, we find that metallicity calibrations based on \niiha \ls are more biased by shocks and AGN at high-z than the \oiiihb/\niiha \ls calibration. 
\end{abstract}


\nopagebreak

\section{Introduction} 

Optical and near-infrared surveys over the last 15 years have firmly established that star-forming galaxies (SFGs) at \ztwo have very different morphological structure in their rest-frame UV/optical light than their local counterparts \citep{CowHuSon95,vdBer+96,ElmElmShe04,Elm+09,ElmElm05,ElmElm06,Gen+06,Gen+08,Gen+11,Law+07,Law+09,Law+12,For+09,For+11a,Wuy+12,Swi+12a,Swi+12b,Jon+10,Jon+12}. Although a majority are characterized as rotating disks with kinematic analysis, they exhibit clumpy structure and elevated velocity dispersions, indicating much larger scale heights than local disks \citep{ElmElm06,For+06,For+09,Law+07,Law+09,Gen+08,Gen+11,Wri+07,Wri+09,Sha+08,Cre+09,vSta+08,Epi+09,LemLam10,Jon+12,Wis+12,Swi+12a,Swi+12b,Epi+12,New+13a}. In addition, their star formation rates (SFR), star formation rate surface densities (\sigsfr) and interstellar pressures are more similar to those of local starbursting galaxies than local normal star-forming disks \citep{New+12a}. 

Given the differing physical conditions in local and high-z SFGs, one might expect this to be reflected in their nebular excitation properties. Indeed, high-z SFGs often fall in the `composite' region of the BPT \citep{BPT81} diagram (the plane of \oiiihb \ls versus \niiha), between HII regions and AGN, coincident with the location of many local starburst galaxies \citep{Tec+04,Kew+06,Erb+06a,Liu+08,Tru+13,Jun+11,Kri+07,Hay+09,Yab+12,Yua+13}. In fact, \cite{Liu+08} and \cite{Bri+08} have found among local SDSS galaxies a positive correlation between relative offset from the main star-forming HII branch of the BPT diagram toward the `composite' region and ionization parameter, SFR, and electron density. They have suggested that higher values of these galaxy parameters could explain the position of high-z galaxies in the BPT diagram as well. Conversely, the offset of \ztwo galaxies from the star-forming branch of the BPT diagram could also be due to the presence of AGN or shocks from large-scale outflows. Unfortunately, weak AGN can be difficult to identify in integrated line ratios, but spatially-resolved data can help to uncover them as demonstrated by the case study of a z = 1.6 galaxy by \cite{Wri+09,Wri+10}. More recently, \cite{New+12a} found that for two star-forming clumps in a \ztwo SFG, elevated \niiha \ls and \siiha \ls ratios were at least in part due to shocks from a galactic-scale outflow originating from the clumps, with no evidence for an underlying AGN. Determining the cause(s) of the offset of high-z galaxies in the BPT diagram is instrumental for understanding the physical properties of these objects and in particular the contribution of AGN to their evolution.

In addition to understanding the physical properties of these galaxies, quantifying the effects of shocks and AGN on emission line ratios will help us to better estimate their gas phase metallicity. These metallicities are essential for properly understanding the interplay between gas inflow, outflow and star formation at the peak epoch of the star formation rate density at z $\sim$ 2--3 \citep{Lil+13}. Rest-optical nebular line ratios calibrated locally are the main technique used for measuring metallicity at high-z. However, different diagnostics can give very different results \citep[see e.g.][]{KewEll08}, and if the \niiha \ls ratio is indeed elevated in many high-z galaxies by either the physical conditions in HII regions, the presence of shocks or an AGN, metallicities determined using this ratio could be systematically overestimated. 

In this work, we present K-, H- and/or J-band datasets for 22 z $\sim$ 1.4--2.5 rest-UV/optically selected SFGs obtained as part of the SINS/zC-SINF and LUCI surveys \citep{For+09,Man+11,Kur+13}. Thus for each galaxy, we measure \ha, \nii, \oiii, and \hb \ls fluxes. The faintness of these lines and the requirement that all, or a significant subset of them should be observable between the bright night sky OH lines in the near-IR atmospheric windows make such studies from the ground very challenging. The SINS/zC-SINF and LUCI surveys allowed us to cull a sizeable amount of suitable targets for a first exploration of spatially-resolved metallicities and excitation in massive \ztwo SFGs. We adopt a $\Lambda$CDM cosmology with $\Omega_{m}$=0.27, $\Omega_{b}$=0.046 and H$_{0}$=70 km/s/Mpc \citep{Kom+11}, as well as a \cite{Cha03} initial stellar mass function (IMF). 

\section{Observations and Analysis}

\subsection{Source selection, observations and data reduction of SINS/zC-SINF galaxies}
Half of our sample (10 SFGs) was drawn from the SINS and zC-SINF surveys of IFU spectroscopic data of z $\sim$ \ls 1.5--2.5 SFGs, obtained using SINFONI on the Very Large Telescope (VLT) of the European Southern Observatory (ESO). For all of these objects, we obtained either K- and H-band (for z = 2--2.5) or H- and J-band (for z = 1.4) data in the seeing-limited mode of SINFONI. For a subset of these objects (six), we also obtained K-band data in the AO mode with the aid of either a natural or laser guide star. The seeing-limited data have a pixel scale of 0.125'' $\times$ 0.25'', whereas the AO data have a pixel scale of 0.05'' $\times$ 0.1'', with a spectral resolution of R = 1900, 2900 and 4500 (5000) for J, H and K-band (AO), respectively.  

The full SINS and zC-SINF samples were selected as described in \cite{For+09} and \cite{Man+11}. Of the final 10 galaxies used in this analysis, two were selected based on their \textit{K}-band magnitudes \citep{Dad+04,Mig+05}, three based on their \textit{K}-band magnitudes and BzK colors \citep{Kon+06,Lil+07}, and five from their U$_{n}$GR colors \citep{Erb+03,Ade+04,Ste+04,Erb+06,Law+09}.

We further selected galaxies to be observed in H- or J-band based on the requirement that they fall at a redshift such that all four emission lines of interest (\nii, \ha, \oiii, \hb) are unaffected or weakly affected by OH sky lines. In addition, these targets are among the brighter objects in the SINS sample, enabling us to observe the relatively weak \hb \ls and \oiii \ls lines in comparatively short exposure times. For the present study, we included only galaxies that had a secure detection in all 4 emission lines with minimal interference from an OH sky line or had a secure detection in 3 lines and an upper limit for one of the lines, with the requirement that there was not an OH sky line near the upper limit, in order to ensure a meaningful value. Out of 15 galaxies for which we attempted to measure all four lines, only 10 were deemed acceptable based on the aforementioned criteria. 

The resulting 10 targets present an exemplary sample of the galaxies in the SINS survey. They span the range in stellar mass from 0.77--31.6 $\times 10^{10}$ \msun, and star formation rate from 25 to 340 \msun/yr. They include various kinematic types: rotation and dispersion dominated objects as well as merging systems. For all the targets, the \ha \ls  and \nii \ls data had already been taken during previous SINFONI observing runs. We note that two of the SINS targets from our sample (Deep3a-15504 and K20-ID5) have evidence for an AGN from X-ray, UV, and/or mid-IR data \citep{For+09,For+13a}. 

Table 1 lists all observing runs during which the data of the 10 sources with secure \ha, 
\nii, \oiii, and \hb detections were obtained (observing details for other sources with only \ha+\nii \ls data can be found in \cite{For+09}). The table also gives the band/grating, pixel scale, observing mode, and the total on-source integration time. For simplicity `125' refers to the largest scale with nominal pixels of 0.125''$\times$0.250''and field of view of 8''x8'', and `50' refers to the intermediate scale with nominal pixels of 0.05''$\times$0.125'' and a field of view of 3.2''x3.2''. 

\begin{table}
\caption{Observation Details for SINS/zC-SINF Targets}
\scriptsize{
\renewcommand{\tabcolsep}{2pt}

\begin{tabular*}{0.5\textwidth}{l l r r r l}
\hline
\hline
\textbf{Object} & \textbf{Band} & \textbf{Scale} & \textbf{Mode} & \textbf{t$_{int}$} & \textbf{Observing Date(s)} \\
 & & [mas] & & [s] & [MonthYY] \\
 \hline 
 \\
 Q1307-BM1163 & J & 125 & seeing-limited & 7200 & Mar05 \\
  & H & 125 & seeing-limited & 14400 & Mar05 \\
 Q2343-BX389 & H & 125 & seeing-limited & 15900 & Jun06, Aug06 \\
 & K & 125 & seeing-limited & 14400 & Oct05 \\
 & K & 50 & LGS-AO & 18000 & Sep12, Oct12 \\
 Q2343-BX610$^{a}$  & H & 125 & seeing-limited & 30000 & Oct05, Nov07 \\ 
 & K & 125 & seeing-limited & 10800 & Jun05, Aug05 \\
 & K & 50 & LGS-AO & 30000 & Sep11, Oct11, Aug12 \\
Deep3a-6004$^{a}$ & H & 125 & seeing-limited & 13200 & Feb10, Mar10 \\
 & K & 125 & seeing-limited & 36000 & Mar06, Mar07 \\
 & K & 50 & LGS-AO & 19200 & Jan10, Mar10, Jan11, \\
 & & & & & Mar11, Mar12, Apr13 \\
Deep3a-15504$^{b}$ & H & 125 & seeing-limited & 14400 & Apr06 \\
 & K & 125 & seeing-limited & 21600 & Mar06 \\
 & K & 50 & LGS-AO & 82800 & Mar06--Apr11 \\
 ZC-407302 & H & 125 & seeing-limited & 20400 & Dec09, Jan10 \\
  & K & 125 & seeing-limited & 7200 & Mar07 \\
  & K & 50 & LGS-AO & 68400 & Apr07, Apr09, Mar12 \\
 K20-ID5$^{b}$ & H & 125 & seeing-limited & 7200 & Mar05 \\
  & K & 125 & seeing-limited & 9600 & Mar05 \\
 K20-ID7 & H & 125 & seeing-limited & 24000 & Jan10, Feb10 \\
  & K & 125 & seeing-limited & 31200 & Oct05, Nov06 \\
 SSA22a-MD41 & H & 125 & seeing-limited & 3600 & Aug06 \\
  & K & 125 & seeing-limited & 25200 & Nov04, Jun05 \\
 Q1623-BX599 & H & 125 & seeing-limited & 12000 & Apr11, Jul11 \\
 & K & 125 & seeing-limited & 5400 & Jul04 \\
 & K & 50 & AO & 7200 & Apr10 \\
\hline

\end{tabular*}}

$^{a}$ Has evidence for an AGN based only on spatially resolved line ratios. \\
$^{b}$ Has evidence for an AGN from X-ray, UV, and/or mid-IR data. 

\end{table}

For the data reduction, we used the software package SPRED and custom routines for optimizing the background/OH airglow subtraction. The point spread function (PSF) full-width at half maximum (FWHM) was measured by fitting a 2D Gaussian profile to the combined images of the PSF calibration star taken throughout the observations of a galaxy. More information on the specifics of the data reduction can be found in \cite{For+09}, \cite{Man+11}, and \cite{For+13b}.

\subsection{Source selection, observations and data reduction of LUCI galaxies}

We also include 12 galaxies observed with the LUCI1 NIR camera and spectrograph on the Large Binocular Telescope (LBT) \citep{Age+10,Sei+10}. LUCI1 allows multi-object spectroscopy with up to 23 laser-cut masks allowing arbitrarily placed and oriented slits having widths down to 0.1'' within a 2.5' $\times$ 4' field \citep{Bus+10}. These galaxies were observed in the H and K-bands for a median integration time of 4 hours each, employing 1''-wide slits. With this slit width and medium resolution gratings, the spectral resolution is 2850 in H band and 2900 or 1900 in K band. The observations were taken between December 2009 and May 2012, with seeing conditions ranging between 0.6'' and 1.1''. For further description of the LUCI1 instrument and the observations, see \cite{Bus+10} and \cite{Kur+13}.

The majority of LUCI targets were selected from the GOODS-N field based on the PEP multiwavelength catalog \citep{Ber+11} with spectroscopic redshifts from \cite{Bar+08}. In addition, galaxies were selected from the Q2343 field \citep{Ste+04} based on the rest-UV selected sample presented in \cite{Erb+06}. Objects from both fields were selected based on their optical redshifts such that there was minimal overlap of the \ha, \nii, \oiii, and \hb \ls emission lines with OH sky features.

The data were reduced using a custom pipeline developed at MPE \citep{Kur+13}. The pipeline includes cosmic ray removal, correction for distortion and the same routine to remove OH sky features as used for the SINFONI reduction \citep{Dav+07}. The background is subtracted from frames in pairs obtained from on-slit dithering. Wavelength calibration was conducted using OH sky lines, and flux calibration using a Telluric standard star. The noise spectrum was determined using the same method as for the SINFONI data \citep{For+09}.

\subsection{Galaxy Properties for the Total sample} 
Our total sample of 22 galaxies contains 10 SFGs from the SINS/zC-SINF survey with all four emission lines detected, 6 galaxies from the LUCI survey with all four lines detected and 6 galaxies from the LUCI survey with an upper limit for either the \nii \ls or \oiii \ls emission line and secure detections for the remaining three lines. These galaxies sample the z $\sim$ \ls 1.5--2.5 `main sequence' of SFGs in the stellar mass/star-formation rate plane between stellar masses of 1$\times$10$^{9}$ and 3$\times$10$^{11}$ \msun \ls and SFRs between 6 and 340 \msun/yr (Figure 1).

For the SINS/zC-SINF galaxies, stellar masses and star-formation rates were determined from stellar population synthesis modeling of broad-band photometry in \cite{For+09}, \cite{Man+11}, and \cite{For+11a}, and assume either a constant star-formation history or an exponentially declining model with \cite{BruCha03} stellar evolution tracks. For Q1307-BM1163, photometry is only available in three optical bands, preventing reliable SED modeling to derive the stellar properties. Galaxy properties for the LUCI sample were derived by \cite{Wuy+11b} based on broad-band SEDs. Stellar masses and star-formation rates are computed with the FAST routine \citep{Kri+09} using \cite{BruCha03} stellar evolution models with exponentially declining star-formation histories. These galaxy properties are listed in Table 2.

\begin{figure}
\centerline{
\includegraphics[width=3.5in]{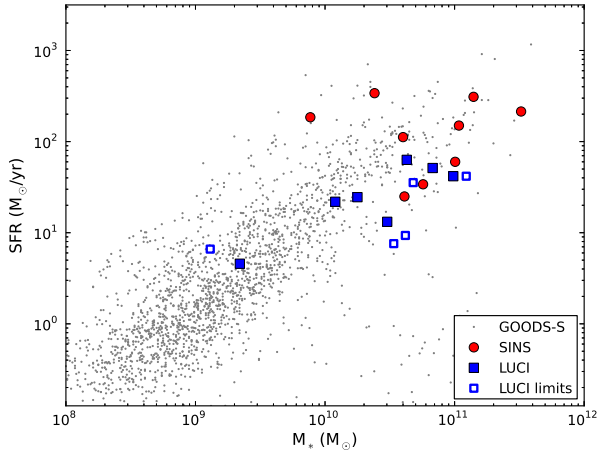}}
\caption{The location of our sample on the SFR-\mstar \ls plane. GOODS-S galaxies from z=1.4--3 are shown as gray points \citep[see:][]{Wuy+12}, red circles are SINS galaxies and blue squares are LUCI galaxies. For the LUCI galaxies, we show objects that have secure detections in all four emission lines with closed symbols and those that have an upper limit in one as open symbols.}
\end{figure}

\begin{table}
\caption{Galaxy properties for the total sample}
\begin{center}
\begin{tabular}{l c c c c c}
\hline
\hline
\textbf{Object} & \textbf{SFR} &  \textbf{sSFR} & \textbf{\mstar} \\
   & \msunyr &  Gyr$^{-1}$ & 10$^{11}$ \msun  \\
 \hline 
 Q1307-BM1163$^{a}$ & 57 & --- & ---  \\
 Q2343-BX389 & 25 & 0.61 & 0.41  \\
 Q2343-BX610$^{b}$ & 60 & 0.60 & 1.01  \\
 Deep3a-6004$^{b}$ & 214 & 0.68 & 3.26  \\
 Deep3a-15504$^{c}$ & 150 & 1.38 & 1.08  \\
 ZC-407302 & 340 & 13.93 & 0.241  \\
 K20-ID5$^{c}$ & 310 & 10.11 & 1.40  \\
 K20-ID7 & 112 & 2.84 & 0.40 \\
 SSA22a-MD41 & 185 & 24.03 & 0.077  \\
 Q1623-BX599 & 34 & 0.60 & 0.571  \\
\hline
GN-7573 & 8 & 0.22 & 0.339  \\
GN-21496 & 9 & 0.22 & 0.417  \\
GN-16564 & 7 & 5.08 & 0.013 \\
GN-31720 & 35 & 0.74 & 0.479  \\
GN-4126 & 23 & 0.60 & 0.380  \\
Q2343-BX442 & 42 & 0.34 & 1.230  \\
GN-2602 & 63 & 1.48 & 0.427  \\
GN-3493 & 51 & 0.76 & 0.676 \\
GN-4104 & 8 & 1.45 & 0.056 \\
GN-28625 & 17 & 1.20 & 0.138 \\
GN-12976 & 6 & 2.57 & 0.023  \\
Q2343-BX436 & 54 & 11.75 & 0.046  \\
\hline

\end{tabular}
\end{center}
$^{a}$ This galaxy does not have reliable photometry, and the SFR is thus derived from \ha. \\
$^{b}$ Has evidence for an AGN based only on spatially resolved line ratios. \\
$^{c}$ Has evidence for an AGN from X-ray, UV, and/or mid-IR data. 

\end{table}

\subsection{Extraction of Emission Line Fluxes and Maps}
We extracted source- and region-integrated emission line fluxes as well as emission line maps from the final SINFONI science data cubes using line profile fitting. Prior to fitting, we apply a 3 pixel wide median filter to the data cubes to increase the signal-to-noise ratio (SNR). We then use the Gaussian fitting routine LINEFIT \citep{Dav+11} to produce \ha \ls emission line flux, velocity and velocity dispersion maps. Using these \ha \ls kinematic maps, we velocity-shifted all of the data cubes for each object such that the \ha \ls line falls at the same wavelength for each pixel in order to remove the line-broadening effect of large scale velocity gradients. We then summed the spectra of individual pixels from the velocity-shifted cubes to obtain co-added spectra. For those galaxies that have high enough SNR data to create kinematic maps for \oiii \ls in addition to \ha, we find that the kinematics of these lines are roughly consistent, such that we are justified in using the kinematic maps of \ha \ls to correct for all of the data cubes. We note that the emission line ratios do not vary significantly without this velocity shifting, but this technique allows us to recover additional flux, which is especially important for some of the fainter lines.

We obtain fluxes for each emission line by simultaneously fitting a constant continuum offset and the \ha, \nii$\lambda$6548,6584, and \sii$\lambda$6717,6731 lines of the co-added spectra assuming identical kinematics (velocity and line width). We then fit the \oiii$\lambda$4959,5007 and \hb \ls lines assuming the same kinematics as for \ha. For all lines, we assume a Gaussian shape. The errors for the line fluxes are generated from the fit uncertainties, using the noise cube produced from the reduction as input. These errors are on average greater than those obtained from 1000 Monte Carlo simulations. 

In order to compare region-integrated spectra or emission line maps (for pixel-to-pixel analysis) that were derived from different bands, we used the 250mas data cube for both even when a 100mas cube was available, for resolution consistency. However, we note that for the integrated spectra, we used the higher quality (from longer t$_{int}$) 100mas cubes when available. For comparing emission lines from different data cubes, we register the position of the galaxy in each cube using the centroid of the continuum emission, if possible. If there is not a clear continuum signature free from residuals, we use the centroid of the \ha \ls and \oiii \ls emission line fluxes, checking that the morphology and peak of emission is similar for these lines. We also find that the point spread functions (PSFs) for different bands are consistent within 30\%, which is the combined uncertainty of the resolution of the data sets. For the pixel-to-pixel analysis, we only consider pixels with a SNR of the line flux \textgreater 3 for all four emission lines. 

For the LUCI targets, we collapsed the 2D data onto a 1D spectrum with an aperture chosen to optimize the SNR. We have attempted a similar velocity-shifting technique as for the SINFONI data, but for the few objects for which this was possible, the procedure only increased contamination from OH features. Thus we have not used any velocity-shifted LUCI spectra, but note that as mentioned before, this should not have a large affect on the emission line ratios.

The emission line fluxes for the LUCI spectra were fit using the LINEFIT routine, with errors derived from 100 Monte Carlo simulations. For lines that were undetected, a 3$\sigma$ upper limit was derived based on the noise spectrum for the wavelengths corresponding to the line of interest based on the \ha-derived redshift, assuming the same line width as \ha. For two of the galaxies from the LUCI sample (GN2602, GN3493), the \oiii $\lambda$5007 line was out of the range of the filter and thus the \oiii \ls upper limit was determined from the \oiii $\lambda$4959 emission line and assuming the scaling \oiii $\lambda$5007 = 3 $\times$ \oiii $\lambda$4959. In another one of our objects (GN4126) the \oiii \ls and \hb \ls fluxes are derived from an HST/WFC3 Grism spectrum (B. Weiner, private communication), because neither of these lines were detected in our LUCI data. These fluxes are determined using the same method as for the SINFONI spectra, except the line positions and widths were allowed to differ from those of \ha \ls since the GRISM line width is more sensitive to the object size than kinematics (because the GRISM spectra are slit-less and very low resolution). Table 3 lists the \nii$\lambda$6584/\ha \ls and \oiii$\lambda$5007/\hb \ls line ratios for our sources.

Galaxy-averaged spectra are shown for the SINS galaxies in Figures 2 and 3 and for the LUCI galaxies in Figures 4 and 5 (the latter of which contains galaxies with an upper limit for one of the lines).

\begin{figure*}[p!]
\centerline{
\includegraphics[width=7in]{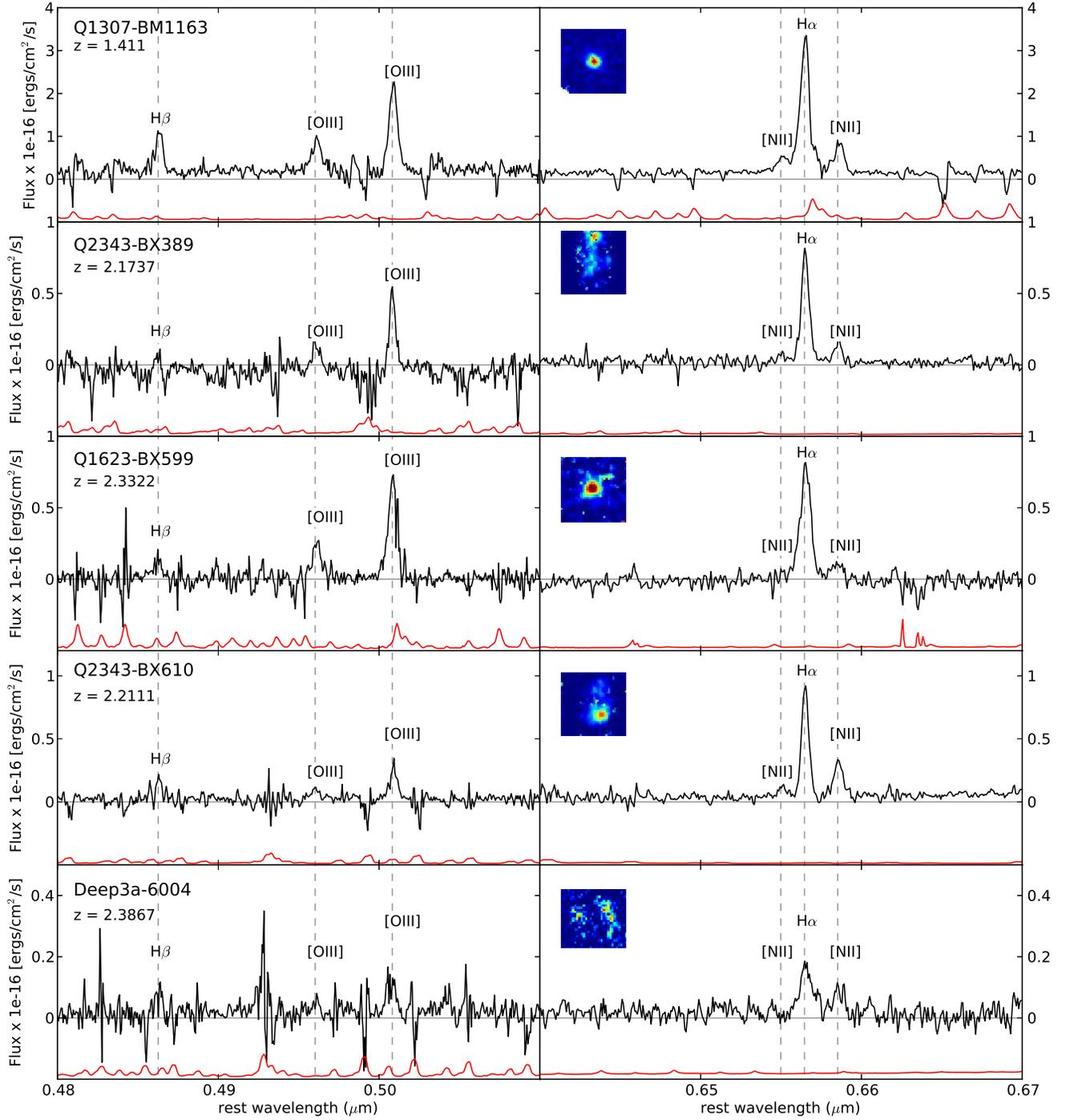}}
\caption{Galaxy spectra for five of the SINS/zC-SINF objects. The black lines show the data and the red lines show the error derived from the noise cubes. The inset in the right panels shows an AO-scale (if available) \ha \ls image of the galaxy. }
\end{figure*}

\begin{figure*}[p!]
\centerline{
\includegraphics[width=7in]{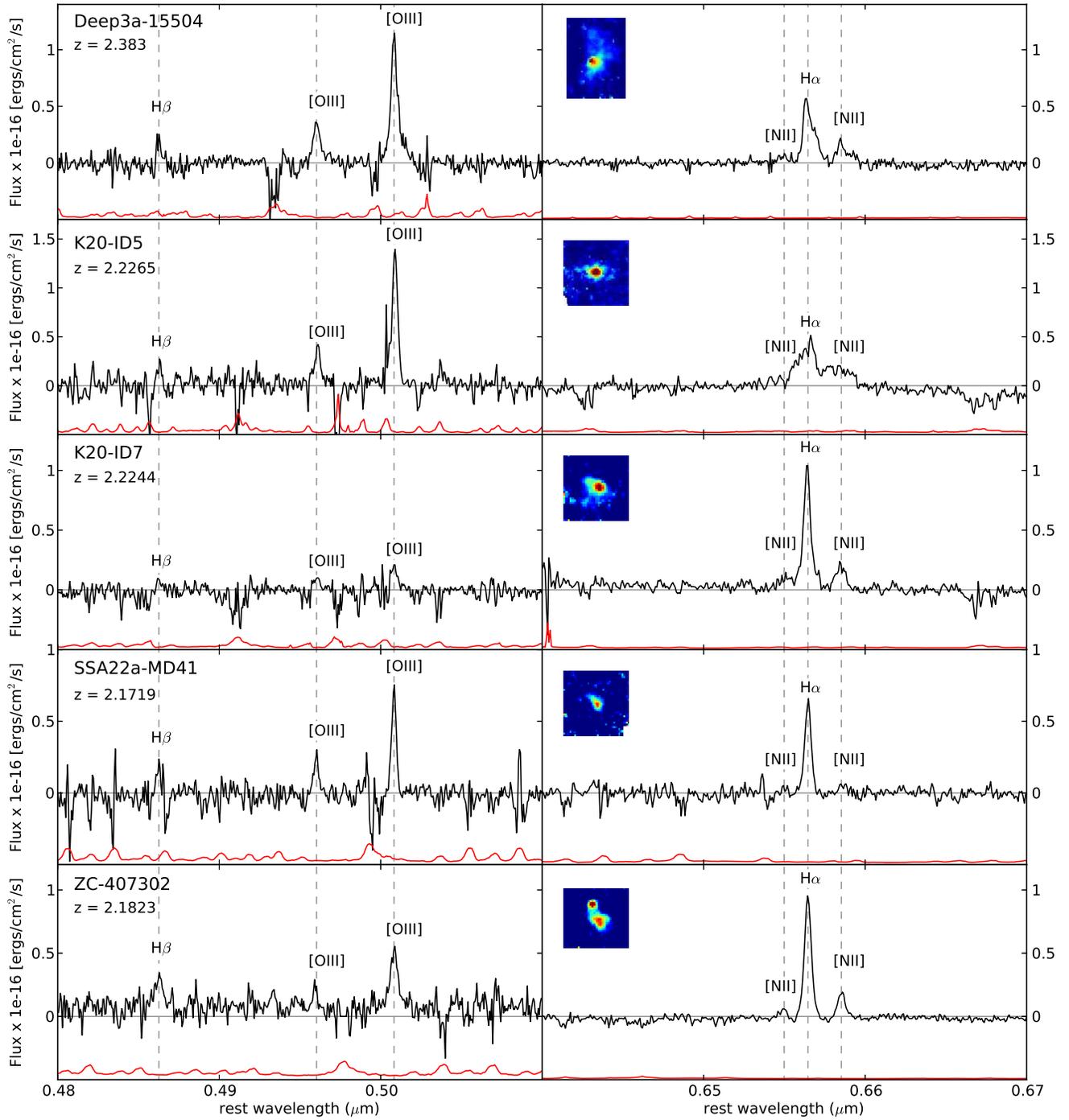}}
\caption{Same as for Figure 2}
\end{figure*}

\begin{figure*}[p!]
\centerline{
\includegraphics[width=7in]{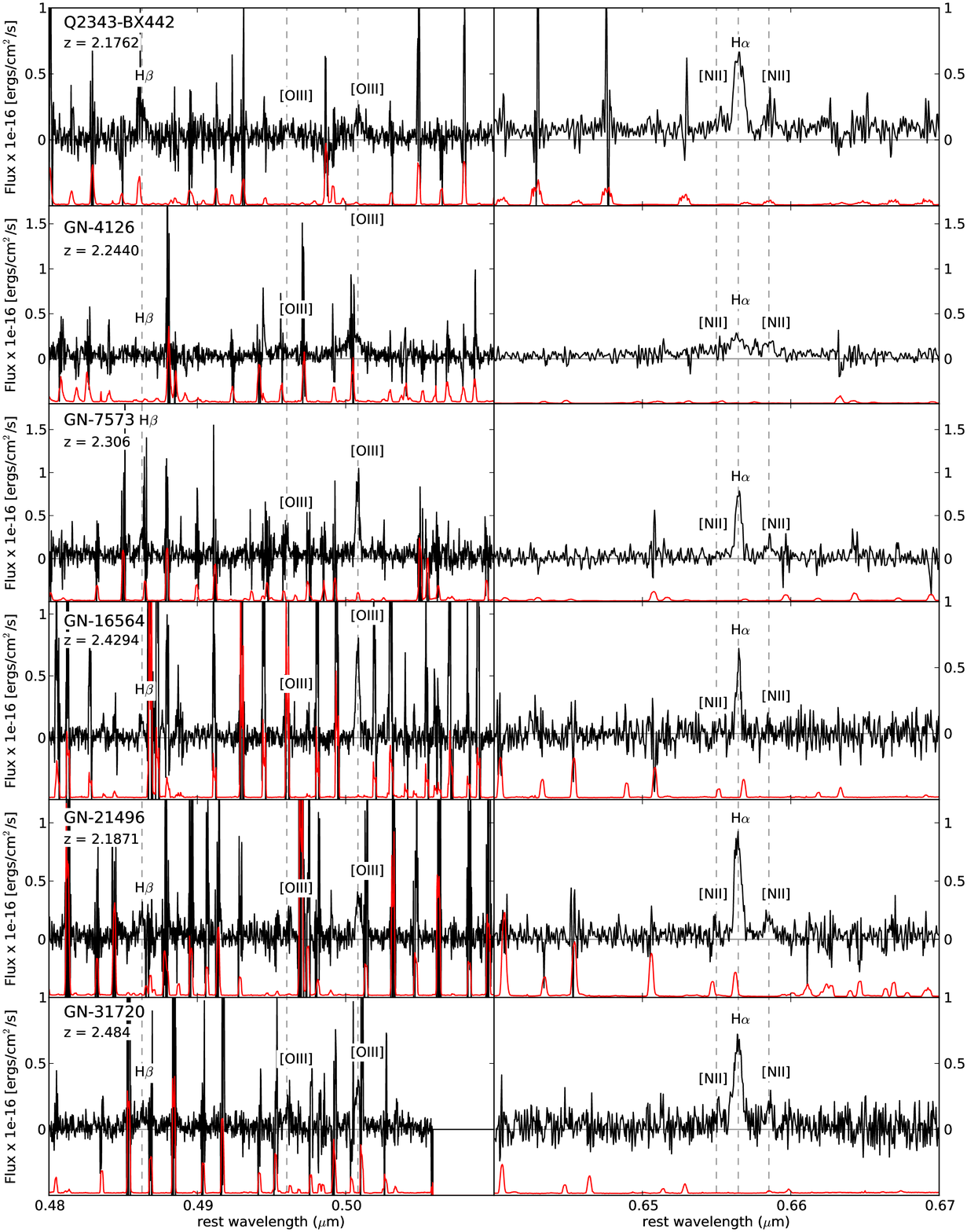}}
\caption{Galaxy spectra of LUCI galaxies with detections in all four lines. The error spectrum (red line) is shown as a factor of 5 lower than it's actual value.}
\end{figure*}

\begin{figure*}[p!]
\centerline{
\includegraphics[width=7in]{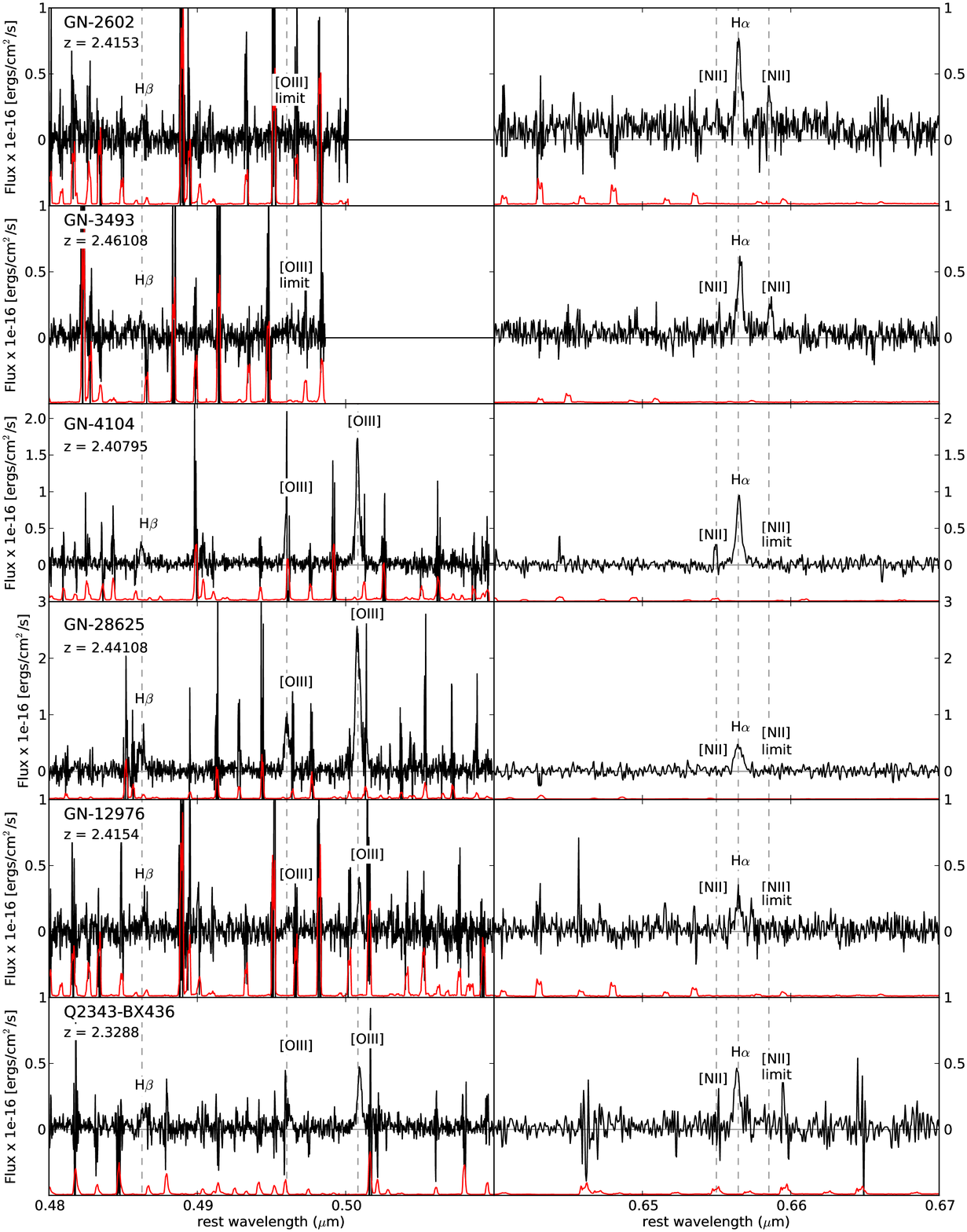}}
\caption{Galaxy spectra of LUCI galaxies with detections in three lines and an upper limit in the remaining line. For GN-2602 and GN-3493, the \oiii \lam 5007 line flux was derived from 3 $\times$ the \oiii \lam 4959 line flux. The error spectrum (red line) is shown as a factor of 5 lower than it's actual value.}
\end{figure*}

\begin{table*}[t]
\caption{Emission Line Ratios}
\begin{center}
\begin{tabular}{l c c c c}
\hline
\hline
\textbf{Object} & \textbf{\niiha} &  \textbf{\oiii $\lambda$5007/\hb} & \textbf{Z/Z$_{\odot}$ (N2)$^{a}$} & \textbf{Z/Z$_{\odot}$ (O3N2)$^{c}$} \\
  \hline 
 Q1307-BM1163 & 0.256 $\pm$ 0.024 & 2.29 $\pm$ 0.16 & 0.75 & 0.45 \\
 Q2343-BX389 & 0.180 $\pm$ 0.027 & 7.04 $\pm$ 1.91& 0.61 & 0.34 \\
 Q2343-BX610$^{b}$ & 0.339 $\pm$ 0.029 & 1.69 $\pm$ 0.22 & 0.88 & 0.66 \\
 Deep3a-6004$^{b}$ & 0.419 $\pm$ 0.227 & 1.62 $\pm$ 0.30 & 0.99 & 0.71 \\
 Deep3a-15504$^{c}$ & 0.350 $\pm$ 0.089 & 8.13 $\pm$ 2.00 & 0.89 & 0.40 \\
 ZC-407302 & 0.223 $\pm$ 0.015 & 1.80 $\pm$ 0.18 & 0.69 & 0.56 \\
 K20-ID5$^{c}$ & 0.560 $\pm$ 0.102 & 7.00 $\pm$ 0.78 & 1.17 & 0.49 \\
 K20-ID7 & 0.236 $\pm$ 0.042 & 2.76 $\pm$ 0.58 & 0.71 & 0.50 \\
 SSA22a-MD41 & 0.100 $\pm$ 0.038 & 4.09 $\pm$ 0.71 & 0.44 & 0.33 \\
 Q1623-BX599 & 0.167 $\pm$ 0.030 & 6.48 $\pm$ 1.33 & 0.58 & 0.34 \\
\hline
GN-7573 & 0.247 $\pm$ 0.040 & 4.17 $\pm$ 0.69 & 0.73 & 0.44 \\
GN-21496 & 0.204 $\pm$ 0.034 & 2.53 $\pm$ 0.70 & 0.66 & 0.49 \\
GN-16564 & 0.067 $\pm$ 0.055 & 7.13 $\pm$ 2.184 & 0.35 & 0.25 \\
GN-31720 & 0.162 $\pm$ 0.060 & 2.97 $\pm$ 0.86 & 0.58 & 0.43 \\
GN-4126 & 0.626 $\pm$ 0.047 & 7.10 $\pm$ 2.28 & 1.24 & 0.50 \\
Q2343-BX442 & 0.232 $\pm$ 0.076 & 0.56 $\pm$ 0.10 & 0.71 & 0.83 \\
GN-2602 & 0.402 $\pm$  0.066 & \textgreater 1.573  $\pm$ 0.72 & 0.96 & 0.71 \\
GN-3493 & 0.143 $\pm$ 0.043 & \textgreater 1.028 $\pm$ 0.79 & 0.54 & 0.58 \\
GN-4104 & \textless 0.109 $\pm$ 0.109 & 6.92 $\pm$ 0.65 & 0.46 & 0.29 \\
GN-28625 & \textless 0.119 $\pm$ 0.119 & 6.30 $\pm$ 1.04 & 0.48 & 0.31 \\
GN-12976 & \textless 0.306 $\pm$ 0.208 & 5.76 $\pm$ 3.50 & 0.83 & 0.43 \\
Q2343-BX436 & \textless 0.151 $\pm$ 0.151 & 4.73 $\pm$ 0.65 & 0.55 & 0.36 \\
\hline

\end{tabular}
\end{center}
$^{a}$ Assuming Z$_{\odot}$ = 8.69 \citep{Asp+09} and using the calibration of \cite{PetPag04}. \\
$^{b}$ Has evidence for an AGN based only on spatially resolved line ratios. \\
$^{c}$ Has evidence for an AGN from X-ray, UV, and/or mid-IR data. 

\end{table*}

\begin{figure*}
\centerline{
\includegraphics[width=7.5in]{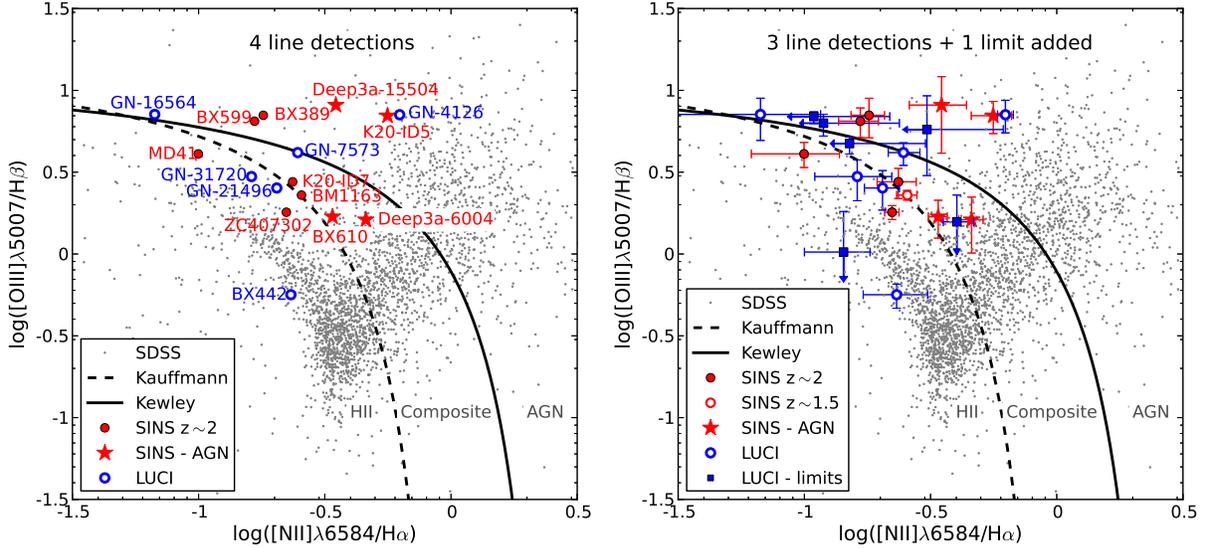}}
\caption{The \niiha \ls vs. \oiiihb \ls plane (BPT diagram). SINS galaxies are shown as either red circles (star-forming) or red stars (those identified as AGN), and LUCI galaxies are shown as blue circles, with galaxies that have only three line detections and one upper limit shown in the right panel as blue filled squares.}
\end{figure*}

\section{Results}
\subsection{Excitation Properties of \ztwo Galaxies}

\subsubsection{Galaxy-integrated measurements}

We show the galaxy-integrated line ratios in the BPT diagram in Figure 6. Many of the objects in our sample are located in the `composite' region of the BPT diagram, between the star-forming and AGN branches traced out by the local SDSS galaxies, and demarcated by the \cite{Kau+03} and \cite{Kew+01a} empirical and theoretical divisions between star-forming and AGN galaxies. This is consistent with most findings of other recent work for samples of galaxies selected by similar criteria \citep[see Figure 7]{Erb+06a,Shapley+05,Liu+08,Kri+07,Hay+09,Yab+12,Tru+13}. The observed offset of high-z galaxies from the local star-forming branch suggests that different physical conditions are prevailing in high-z non-AGN actively star-forming systems compared to the bulk of local star-forming galaxies \citep[see, e.g., the detailed discussion by][]{Liu+08}. We note that all of the galaxies that are identified as AGN either from X-ray, UV or mid-IR data or based on spatially-resolved rest-frame optical data \citep[see next section and][]{For+13a}, have \niiha \ls $\gtrsim$ 0.3 independent of whether they are located in the composite or AGN region of the diagram. However, all galaxies in the composite region are not necessarily AGN. We further examine the galaxies that fall near the composite region and how to identify galaxies with obscured AGN using spatially resolved data in the next section.

\begin{figure}
\centerline{
\includegraphics[width=3.5in]{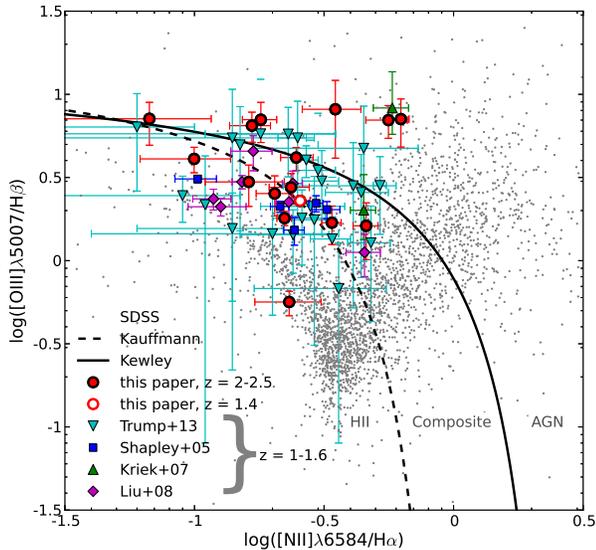}}
\caption{The \niiha \ls vs. \oiiihb \ls plane (BPT diagram). SINS and LUCI galaxies are shown as red circles, and data from \cite{Shapley+05}, \cite{Kri+07}, \cite{Liu+08} and \cite{Tru+13} are blue squares, green upright triangles, magenta diamonds and cyan inverted triangles, respectively. Only secure detections in all four lines (not upper limits) are shown.}
\end{figure}

To explore some of the trends described in \cite{Liu+08} and \cite{Bri+08}, we color-code galaxies in the BPT diagram by SFR, stellar mass and sSFR (Figure 8). We include all of our SINS and LUCI galaxies as well as galaxies from the literature, presented in Figure 7, with secure detections in all four emission lines. Thus this analysis includes galaxies from z=1 to 2.5. As expected from the mass-metallicity relation, there is a trend of increasing stellar mass with higher \niiha \ls ratios. We also find a trend of increasing offset (mostly in the \niiha \ls direction) from the HII branch with increasing SFR, which may be correlated with an increase in ionization parameter, but may also be due to the aforementioned stellar mass trend since stellar mass and star formation rate are tightly correlated for SFGs \citep{Dad+07,Rod+11}. Finally, there may be a (marginal) trend of increasing \oiiihb \ls ratio with increasing sSFR, suggesting that galaxies in the `composite' region of the BPT diagram may be in a burstier phase of star-formation than those that lie on the HII branch. This is consistent with the finding that many of the local galaxies in the `composite' region are starbursts \citep{Liu+08,Bri+08}. Alternatively, the elevated sSFR for these galaxies could be related to AGN activity affecting the line ratio, if there is a correlation between sSFR and AGN activity. However, \cite{Ran+13} suggest that there is no such correlation.

\begin{figure}
\centerline{
\includegraphics[width=3.5in]{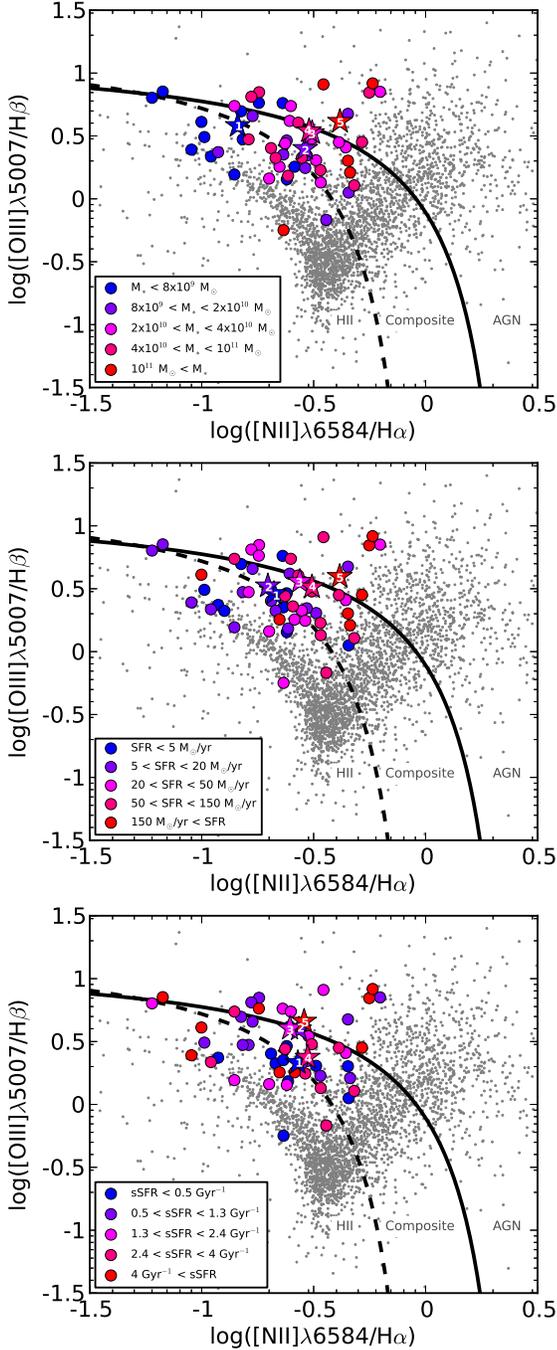}}
\caption{BPT diagram with SINS, LUCI and literature data color-coded by \mstar (top panel), SFR (middle panel) and sSFR (lower panel). Individual galaxies are shown as circles and the average of each bin is shown as a star with a number representing the order of the bin. Both \mstar \ls and SFR appear to increase with increasing \niiha, while there is perhaps a marginal trend of increasing sSFR with increasing \oiiihb.}
\end{figure}

The measured \hb \ls fluxes might also be biased by underlying stellar \hb \ls absorption \citep{Duf+80}. We therefore correct for this absorption using the best fit SED-derived star-formation histories and ages. Once we account for this correction factor, we see a noticeable down-shift in \oiiihb \ls of $\sim$ 0.02 to 0.21 dex (mean 0.08 dex). Nevertheless, accounting for \hb \ls absorption does not reconcile our data with the distribution of local galaxies from the SDSS (Figure 9) and therefore does not affect our qualitative conclusions. Our estimated corrections for \hb \ls absorption are fairly similar to those of \cite{Shapley+05} who reported a maximum down-shift of 0.1 dex and also found that unaccounted-for \hb absorption is likely not causing the observed offset in \oiiihb \ls for their DEEP2 sample. For the remainder of this paper, we use emission line ratios that are not corrected for stellar \hb \ls absorption for consistency with previous work.

\begin{figure}
\centerline{
\includegraphics[width=3.5in]{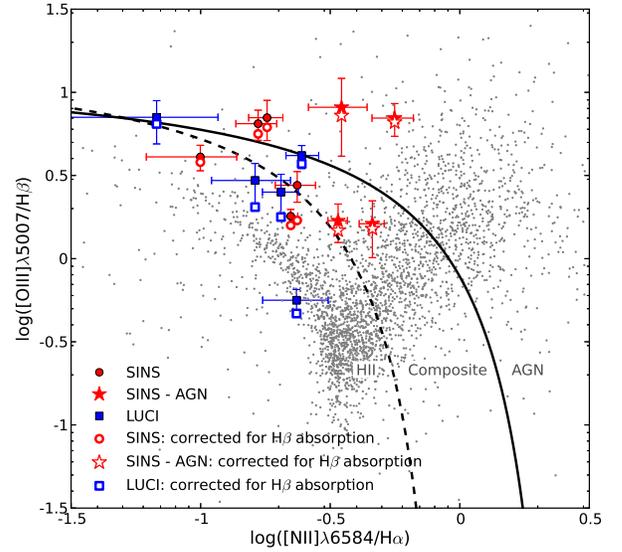}}
\caption{BPT diagram with SINS and LUCI data before (closed symbols) and after (open symbols) correction for stellar \hb \ls absorption. The correction is marginal, with a mean offset of 0.08dex, and cannot fully account for the elevated line ratios of \ztwo SFGs as compared with local SFGs.}
\end{figure}

\subsubsection{Spatially-resolved measurements}

In this section, we exploit the high-quality IFU data taken of the SINS/zC-SINF galaxies in our sample. Overall, the emission line and kinematic maps of our targets for the different emission lines  are in good agreement. For most sources, the main emission peak in all of the individual line maps is found at the same location inside the source boundaries. Some notable exceptions exist for a couple of sources (K20-ID5 and ZC407302, see discussion later in this section), where, in particular, the peak \oiii \ls emission was somewhat offset from the \ha \ls peak emission. For four of the galaxies in our SINS/zC-SINF sample, the SNR is sufficiently high in all four lines (\nii, \ha, \oiii, \hb) that we are able to plot individual pixels in the BPT diagram for a large region of these galaxies. For the remaining galaxies (except Deep3a-6004), we are still able to obtain high quality spectra of the inner (R \textless 0.4'') and outer regions (see Figure 10), where the center is determined as in Section 2.

We find that many galaxies have inner regions that are offset to higher excitation relative to their outer regions, indicating that perhaps the inner region is influenced by an (possibly obscured) AGN and the outer region is dominated by star-formation (see Figure 10). Alternatively, the inner region could have a larger contribution from shock excitation or elevated ionization parameter, perhaps due to an outflow, as seen in the star-forming clumps of a \ztwo SFG presented in \cite{New+12a}. This inner/outer region offset is seen for two galaxies with strong evidence for an AGN, Deep3a-15504 and K20-ID5, as well as Q2343-BX389, which has no other evidence for an AGN (see discussion below). However, we note that the \hb \ls line for Q2343-BX389 becomes somewhat contaminated by an OH sky line on it's redshifted side, and thus its outer spectrum may be unreliable, as reflected in the error bars. For the remaining galaxies, the offsets between the inner and outer regions in the BPT diagram are consistent with the 1$\sigma$ errors. 

\begin{figure}
\centerline{
\includegraphics[width=3.5in]{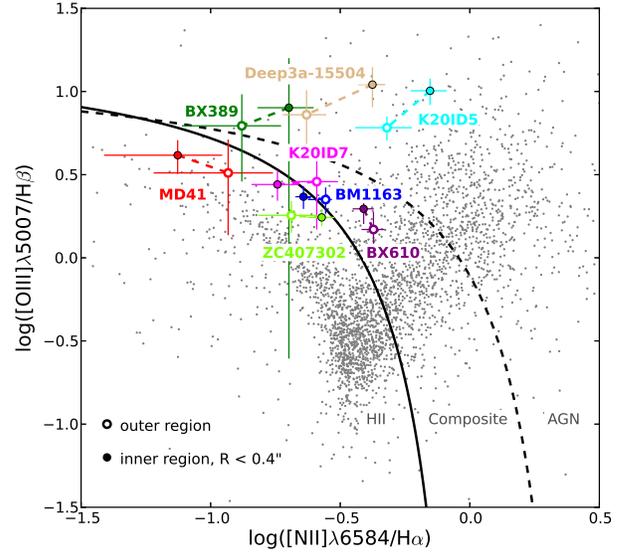}}
\caption{BPT diagram showing inner and outer regions for each galaxy. The inner regions (with R \textless \ls 0.4'') are shown as closed symbols and the outer regions as open symbols. Several objects, particularly those in the AGN regime of the diagram, show elevated line ratios in their inner regions with respect to their outer regions.}
\end{figure}

Figures  11, 12, 13 and 14 show pixel-by-pixel BPT diagrams for Deep3a-15504, ZC407302, Q1623-BX599 and Q2343-BX610 along with \niiha \ls and \oiiihb \ls maps with a SNR cut-off of 3. There are some small discrepancies between Figure 10 and Figures 11 through 14, and these stem from the different techniques used to extract the line ratios in the pixel-by-pixel and inner/outer region analysis and the resulting flux weightings (i.e. fitting the lines to individual pixel spectra versus to integrated spectra). \\

\noindent \textbf{- Deep3a-15504} \\
Deep3a-15504 is a large, rotating massive disk which has previously been identified as an AGN based on broad and/or high excitation lines in the optical (rest-UV) spectrum \citep[E. Daddi, private communication]{Kon+06,Gen+06}. In addition to elevated \niiha \ls values, its nuclear region also demonstrates extremely broad \ha \ls emission line wings, with velocities of up to 1500 \kms \citep{Gen+06,For+13a}. The deep, AO-assisted \ha \ls map obtained for this galaxy shows that the star formation activity in the large disk takes place in several moderately bright clumps superimposed on diffuse emission. With pixel-by-pixel analysis (see Fig. 11), the central AGN-dominated (red and green symbols) and outer star-forming disk regions (blue pixels) separate clearly. While the outer region pixels still lie between the AGN region and the composite region of the BPT diagram, it is clear that the high excitation due to the AGN at the center drives the integrated ratios to values even closer to those of the AGN regime.  \\

\begin{figure}
\centerline{
\includegraphics[width=4in]{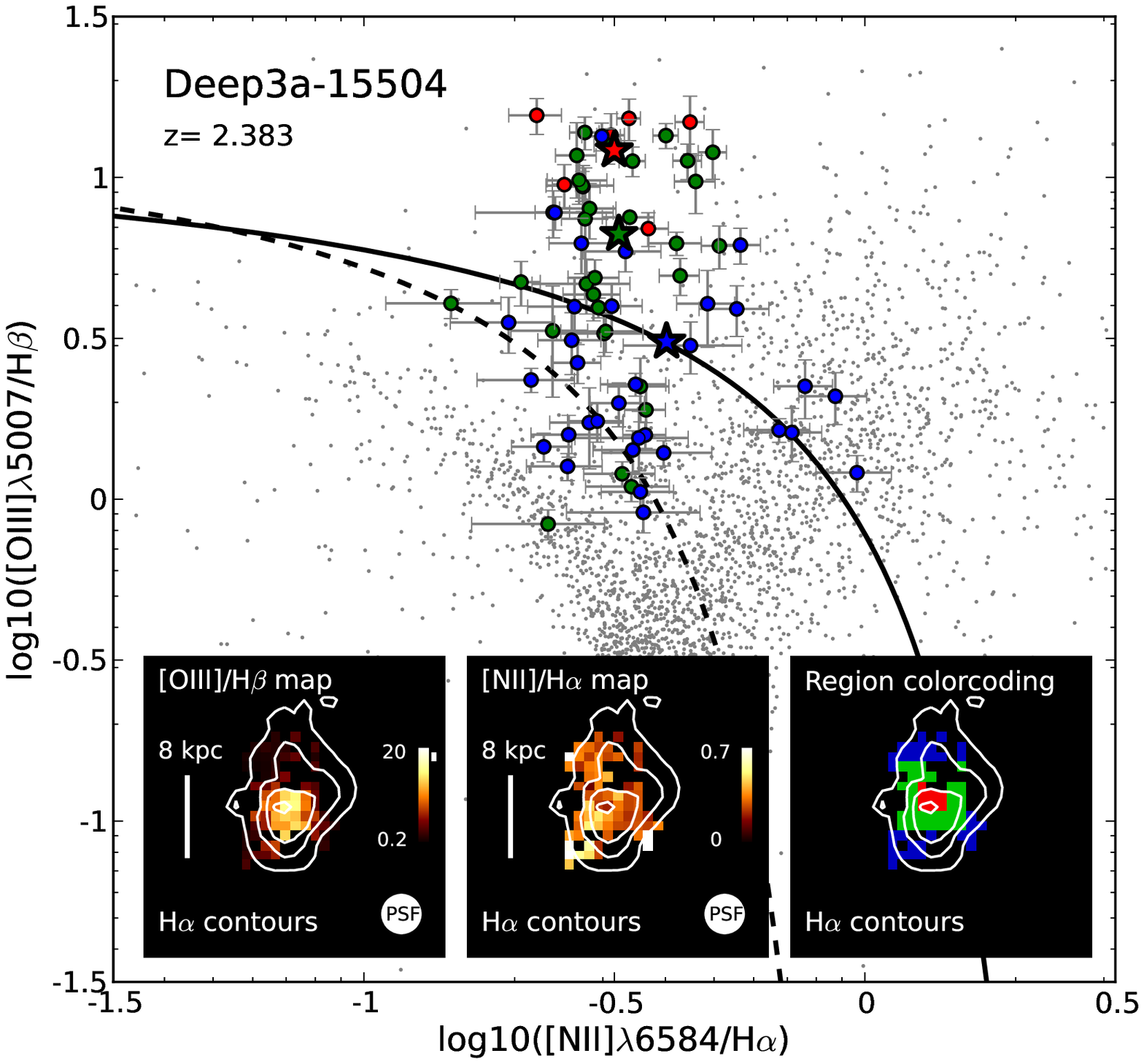}}
\caption{Pixel-by-pixel BPT diagram for Deep3a-15504. Red points are from the innermost region, green points from further out and blue points from the furthest edges of the galaxy. The region-averaged data are shown as star symbols. The rightmost inset shows the color-coding for the regions, the middle inset shows a map of \niiha, and the left inset shows \oiiihb. \ha \ls contours appear on all three maps, and only pixels are shown with SNR \textgreater \ls 3 in all four lines. The emission line maps show higher ratios in the nuclear regions, particularly for \oiiihb \ls and this is reflected in the BPT diagram. Thus the outer region is influenced by star-formation in the disk, while the nuclear region is mostly a reflection of the AGN.}
\end{figure}

\noindent \textbf{- ZC407302} \\
ZC407302 has no evidence for an AGN either from resolved \ha \ls observations or ancillary data. This is corroborated by the pixel-by-pixel BPT diagram (Figure 12), where all pixels are consistent with excitation as observed in purely star-forming objects, and are very similar to the integrated line ratio. This is also reflected in the fairly uniform line ratio maps from the seeing-limited SINFONI data. As mentioned earlier in this section, there is a slight discrepancy between the peaks of \ha \ls and \oiii \ls emission. This could be due to the asymmetric light distribution in \ha \ls and \nii \ls seen in the high resolution AO map \citep[see Figure 3,][]{Gen+08,Gen+11}. The asymmetry is mainly caused by a compact source north-east of the main body of the galaxy, which is not resolved in the seeing-limited maps of Figure 12 but is present in the adaptive optics data set as well as in ACS \textit{i}-band and HST WFC3 J- and H-band imaging available for this galaxy (S. Tacchella \textit{et al.} in preparation). This source could be a bright disk clump at the edge of the system or a second, lower mass galaxy interacting with the main galaxy. The overall velocity field is consistent with a large disk galaxy, while on kpc-scales, deviations from pure rotation are visible and the north-eastern part appears more disturbed \citep{Gen+11}.  \\

\begin{figure}
\centerline{
\includegraphics[width=4in]{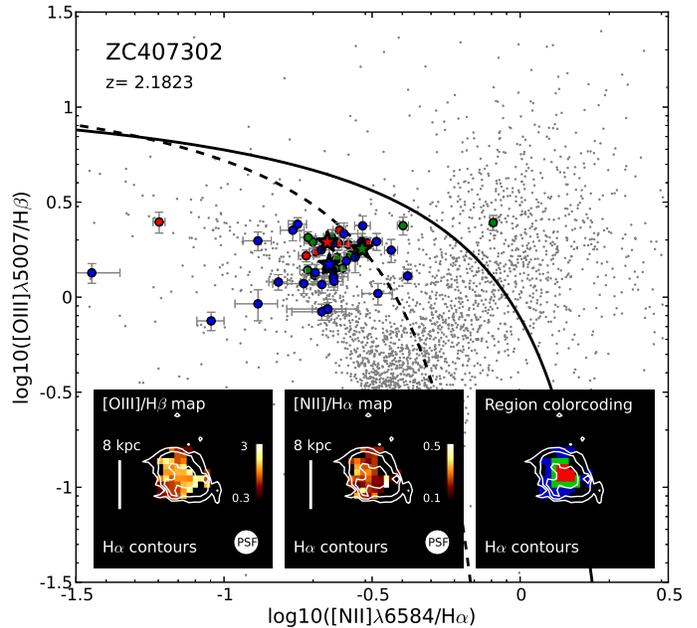}}
\caption{Pixel-by-pixel BPT diagram for ZC407302. Symbols and insets are the same as for Figure 11. The emission line ratio maps are mostly uniform and all regions of the galaxy are consistent with excitation from HII regions.}
\end{figure}

\noindent \textbf{- Q1623-BX599} \\
Q1623-BX599 is a compact, rotating disk \citep{New+13a}. Like ZC407302, it has no evidence for an AGN and all three regions appear to lie in or close to the star-forming region of the BPT diagram (Figure 13).  However, this galaxy does barely lie in the AGN regime of the BPT diagram based on integrated line ratios (Figure 6). This may be due to the fact that much of the east half of the galaxy has very low SNR in \hb \ls (see \oiiihb map in Figure 13) and the inclusion of this region in the integrated spectrum is driving up the \oiiihb \ls ratio. Based on the low \niiha \ls ratio for this galaxy (0.17 $\pm$ 0.03), and the results of the pixel-by-pixel analysis, it is unlikely that this galaxy harbors an AGN. \\

\begin{figure}
\centerline{
\includegraphics[width=4in]{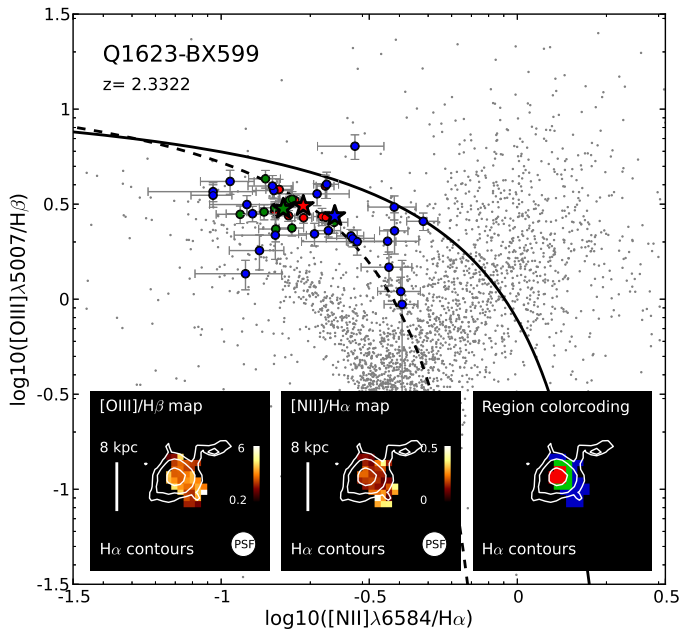}}
\caption{Pixel-by-pixel BPT diagram for Q1623-BX599. Symbols and insets are the same as for Figures 11 and 12. The emission line ratios for all regions are consistent with pure star-formation.}
\end{figure}

\noindent \textbf{- Q2343-BX610} \\
\ha \ls kinematics clearly show that this system is a large rotating disk with several bright clumps visible in the AO SINFONI \ls data as well as in high resolution NIC2 H-band and WFC3 maps \citep[S. Tacchella \textit{et al.} in preparation]{For+11a,For+11b}. While there is no other evidence for an AGN from the available rest-UV spectrum \citep{Erb+06}, there is an indication of a possible AGN from the observed mid-IR IRAC colors \citep{Hai+12,For+11a,For+13a} and resolved analysis of the nuclear region shows elevated emission line ratios (Figure 14, particularly in \oiiihb) and very broad emission line wings \citep[up to 1500 km/s, see:][]{For+13a}, further supporting the presence of an obscured AGN. Interestingly, while the \oiiihb \ls ratio peaks in the same location as the \ha \ls peak, \niiha \ls peaks slightly north of the center, at the continuum- and kinematic-derived center (0.4'' or 3.5 kpc away).

\begin{figure}
\centerline{
\includegraphics[width=4in]{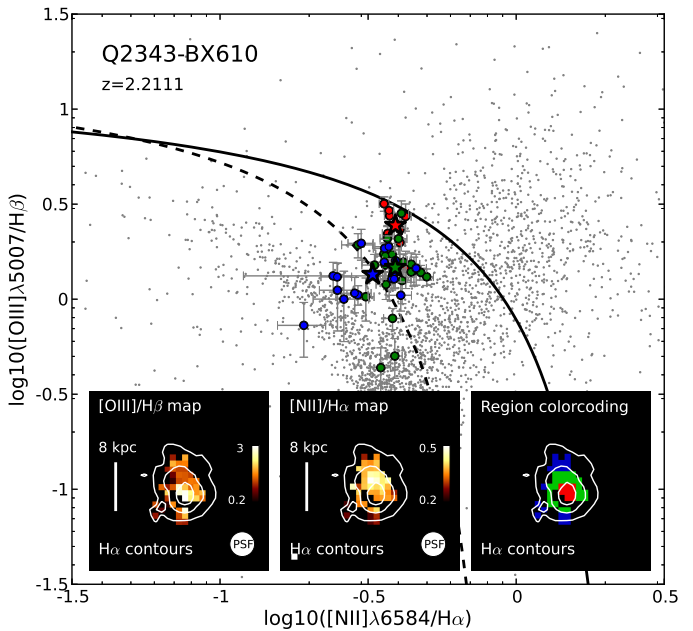}}
\caption{Pixel-by-pixel BPT diagram for Q2343-BX610. Symbols and insets are the same as for Figures 11, 12 and 13. The innermost region has elevated \niiha \ls and \oiiihb \ls ratios, suggestive of an AGN.}
\end{figure}

There are two additional galaxies from our sample (Deep3a-6004 and K20-ID5) that show evidence for an AGN, but do not have a sufficiently high SNR in all four lines for the pixel-by-pixel analysis. Deep3a-6004, with a ring-like \ha \ls morphology and kinematics suggestive of rotation \citep{For+09,New+13a}, is located in the `composite' region of the BPT diagram. While it's VIMOS rest-UV spectrum shows no evidence for an AGN (E. Daddi, private communication), the nuclear region has an elevated \niiha \ls ratio and very broad emission \citep[up to 1500 \kms, see:][]{For+13a}, both suggestive of an AGN. Unfortunately, the \oiii \ls and \hb \ls lines have too low SNR in the nuclear region to be analyzed. K20-ID5, while located in the AGN regime of the BPT diagram, has previously been characterized as a non-AGN galaxy with elevated emission line ratios due only to photoionization and shocks by \cite{vDok+05}. This analysis was based on the optical spectrum, which includes [He{\small II}]$\lambda$1640 and [C{\small II}]$\lambda$1909 emission, but no higher ionization lines such as [C{\small IV}]$\lambda$1549 that would allow unambiguous identification of an AGN \citep{Dad+04}. However, in recent deep 4Ms \textit{Chandra} observations, K20-ID5 has been detected in both hard and soft X-rays \citep{Xue+11}, and it's 1.4 GHz flux density is also consistent with the presence of an AGN. Moreover, it's \textit{Spitzer}/IRAC colors \citep[from the catalog of][]{Wuy+08} also satisfy the criteria for an AGN \citep{Stern+05,Lac+07,Don+12}. These new data support the notion that K20-ID5 does in fact harbor an AGN, though shocks in the nucleus and the disk are not excluded. 

One other galaxy from our sample (Q2343-BX389) lies in the AGN regime of the BPT diagram, although it is not clear if it in fact contains an AGN. First, the \hb \ls absorption correction brings it closer to the `composite' region of the BPT diagram (see Figure 9). Second, the \niiha \ls ratio (0.18 $\pm$ 0.03) is quite low for a galaxy with an AGN. Finally, the \hb \ls line is affected by an OH sky line for a region of the galaxy and this could be affecting its positions in the BPT diagram. On the other hand, Q2343-BX389 is fairly massive (\mstar $\sim$ 4$\times$10$^{10}$ \msun), and close to the stellar mass for which most \ztwo SFGs host an AGN (10$^{11}$ \msun), as seen by \cite{For+13a}, and the inner region of the galaxy is offset towards higher excitation than the outer region (Figure 10). Unfortunately, our current data are insufficient for determining whether or not this galaxy contains an AGN.

\begin{figure*}[t]
\centerline{
\includegraphics[width=7in]{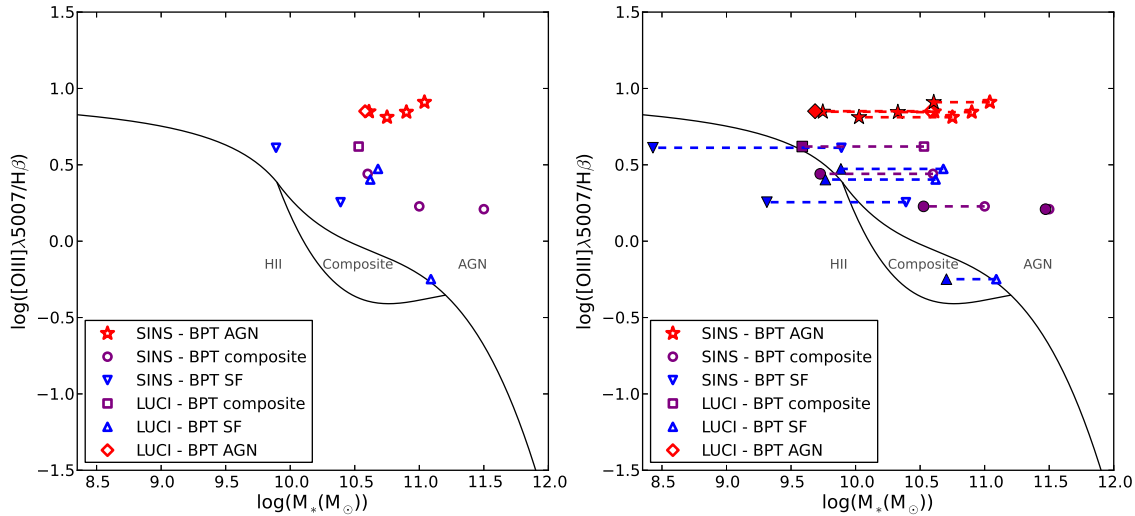}}
\caption{MEx diagnostic from \cite{Jun+11} with SINS/zC-SINF and LUCI galaxies. The left panel shows the traditional MEx diagnostic with the galaxies color and symbol coded by their position in the BPT diagram. Red stars and diamonds are SINS and LUCI AGN, purple circles and squares are SINS and LUCI composite galaxies, and blue inverted and upright triangles are SINS and LUCI star-forming galaxies, according to the BPT diagram. In the right panel, we shift the data points according to the shift in the mass-metallicity relation between z $\sim$ 0 and 2 (using the \cite{Erb+06a} and \cite{Tre+04} relations). The original points are shown as open symbols and the shifted points are shown as closed symbols. While the MEx diagnostic does not accurately predict the location of \ztwo galaxies in the BPT diagram, applying a simple shift according to the mass-metallicity relation brings the two diagnostics into better agreement.}
\end{figure*}

Our results from the analysis of spatially-integrated line ratios combined with other source properties suggest that there is no single dominant mechanism common to all galaxies, which is causing the observed offset in the BPT diagram, compared to the location of local SDSS galaxies.  Known AGNs (Deep3a-15504 and K20-ID5) are located in the AGN region and other sources show a similar offset as found in other high-z studies \citep[see e.g.:][]{Erb+06a,Shapley+05,Liu+08,Kri+07,Hay+09,Yab+12,Tru+13}, being located towards higher values in \niiha \ls and \oiiihb \ls and in between the AGN and star-forming branch. For these objects, contributions from (possibly low-luminosity) AGN (for Q2343-BX610 and Deep3a-6004) as well as (possibly outflow-related) shocks or an ISM with higher electron density and ionization parameter (for Q1307-BM1163, K20-ID7, Q2343-BX389 and Q1623-BX599) may be possible \cite[e.g., ][]{Shapley+05,Liu+08,Erb+06a,Wri+09,Wri+10,New+12a}. We probe the latter two scenarios using photoionization and shock models in section 3.2. 

The key result from this section is that we are able to \textit{identify potential low-luminosity or obscured AGN through the use of resolved line ratio measurements} (e.g. Q2343-BX610), which would not be identified as such using integrated data. Using this technique, we also find that not all high-z galaxies in or near the composite region appear to contain AGN.

\subsubsection{MEx diagnostic at high-z}

Recently, \cite{Jun+11} introduced a new excitation diagnostic that could be used in the absence of an \nii \ls and \ha \ls measurement, the Mass-Excitation diagnostic (MEx). This diagnostic uses the well-known correlation between galaxy stellar mass and gas phase metallicity to create an alternate BPT-like diagram with \oiiihb \ls vs. stellar mass. While both \cite{Jun+13} and \cite{Tru+13} have shown that this relation, which was calibrated locally, holds up to z $\lesssim$ 1.6, separating AGN and star-forming galaxies, we hope to test whether this relation could also be used for our \ztwo objects.

As seen in the left panel of Figure 15, we find that this diagnostic does not work for our \ztwo galaxies. All of our galaxies with all four emission lines (\nii, \ha, \oiii, \hb) fall in the AGN regime of the MEx diagram regardless of their position in the original BPT diagram. This is not surprising given the strong offset of \ztwo galaxies in the stellar mass-metallicity relation. In order to quantify the effect of the evolution of the mass-metallicity relation on this discrepancy, we derive a simple correction for the stellar mass (i.e. what would the stellar mass of our \ztwo galaxies be at z $\sim$ 0 for the same value of \niiha). First, we calculate the shift in stellar mass as a function of metallicity between the \cite{Erb+06a} \ztwo relation and the \cite{Tre+04} SDSS local relation (using the N2 indicator). Then we use the \cite{Erb+06} mass-metallicity relation to get the mass shift as a function of stellar mass and apply this shift to our \ztwo galaxies in Figure 15 (right panel). We note that this correction can not account for the presence of shocks and elevated ionization parameter, which could also affect the MEx diagnostic at \ztwo. 

We find that this corrected MEx diagnostic does a reasonably good job of predicting the location of the galaxies in the BPT diagram: The BPT star-forming galaxies (left of the \cite{Kau+03} relation) all fall in the MEx star-forming regime or close to it; all but one of the BPT composite galaxies are shifted towards the border between the AGN and star-forming regimes; and the BPT AGN galaxies are still in the MEx AGN regime. In fact, we find that the two BPT composite galaxies that still lie in the AGN regime of the `corrected' MEx diagram (Deep3a-6004 and Q2343-BX610) are also classified as AGN when using spatially resolved analysis.

\subsection{Comparison to Photoionization and Shock models}

We generate both photoionization and shock models to explore the effect of these processes on the location of high-z galaxies in the BPT diagram. The photoionization models are created using SEDs generated with Starburst99 \citep{Lei+99,VazLei05}, along with the photoionization code Cloudy \citep{Fer+13}. We assume a spherical geometry, a uniform electron density of 100 cm$^{-3}$, and an ISM dust model. In generating the SEDs we assume a 10 Myr old cloud with a constant SFR of 100 \msunyr (the median for the SINS galaxies), a \cite{Kro01} IMF, high mass-loss stellar evolution tracks from the Geneva group \citep{Mey+94} and a Pauldrach/Hillier atmosphere \citep{Pau+01,HilMil98}. Models are generated for a range of metallicities (for 0.2Z$_{\odot}$, 0.4Z$_{\odot}$ and Z$_{\odot}$) and ionization parameters (logU = -3.5 to -1.8). Considering an open geometry has little effect on the resulting line ratios, as does altering the filling factor (we consider a filling factor of 1 here), changing the density (between 1 and 500 cm$^{-3}$), adding molecules, removing the grain model, or changing the SFR by an order of magnitude.

The shock models are generated according to the method outlined in \cite{GnaSte09} for the same metallicities as above (0.2--1 Z$_{\odot}$) and for shock velocities of 500 and 1000 \kms \citep[comparable to the velocities of the warm ionized component and hot wind fluid in galactic outflows, see][]{Vei+05}. In these shock models, we have calculated the non-equilibrium ionization and cooling, followed the radiative transfer of the shock self-radiation through the postshock cooling layers, took into account the resulting photoionization and heating, and followed the dynamics of the cooling gas. We have also followed the emission-line intensities of several lines produced in the postshock cooling layers.

To obtain models with both photoionization and shocks, we normalize the lines fluxes in each of the photoionization and shock models by the \ha \ls flux, and combine them with varying contributions, as shown in Figure 16. The SINS and LUCI data (particularly in the `composite' region of the BPT diagram) are best fit by the models with a 25\% flux contribution to the \ha \ls emission from shocks, and are fit equally well with either the 500 or the 1000 \kms \ls shock velocity models. Incidentally, this is consistent with what was found by \cite{New+12a} for the fraction of \ha \ls emission deriving from shocks (15--30\%) in winds from individual star-forming clumps in a \ztwo SFG. However, we note that if the \hb \ls absorption correction shown in Figure 9 is applied, the data points may be more consistent with the 100\% photoionization model. Both the 100\% and the 75\% photoionization models imply ionization parameters of logU = -3 to -1.8 for our galaxies. 

If we compare the metallicities from the models with those assumed using the \cite{PetPag04} \niiha \ls calibration (indicated by the grey lines in Figure 16), we find that galaxies determined to have close to solar metallicity based solely on their \niiha \ls ratios, could indeed have metallicities as low as 0.2Z$_{\odot}$ when accounting for photoionization effects and shocks (for both galaxies with and without evidence for AGN). The discrepancy is lower when the metallicities are calibrated using the O3N2 diagnostic (see next section, and Table 3), but there is still a significant offset between that calibration and the metallicities determined from the models. When considering the non-AGN galaxies that fall within the best fit model grid (Q1307-BM1163, K20-ID7, GN-21496), the metallicities (Z/Z$_{\odot}$) calculated with the N2 diagnostic are overestimated (as compared with the models) by a factor of $\sim$3, and those calculated with the O3N2 diagnostic are overestimated by a factor of $\sim$2.

Our models are roughly consistent with those of previous work on photoionization and shock models \citep[see e.g.][]{DopSut95,Dop+00,Dop+06,All+08,Lev+10,Ric+11}. The 100\% photoionization models occupy a similar region of parameter space (mostly in the HII region and part of the composite region), as those from \cite{Dop+00}, \cite{Dop+06} and \cite{Lev+10}, and the shock model produces similar line intensities as for other fast shock codes \citep[e.g.][]{All+08}. We compare our combined shock and photoionization models to those of \cite{Ric+11}, and find good overall agreement, with \niiha \ls increasing with shock fraction, although we find an overall decrease in \oiiihb \ls with shock fraction while they find an increase. This discrepancy may be due to the fact that they use slow shocks with velocities ranging from 100 to 200 \kms, while our shocks are for 500 and 1000 \kms.

From our models, we find that most or all of the galaxies located in the `composite' region of the BPT diagram need not contain an AGN and may only lie there because of shocks and photoionization (particularly if we correct \hb \ls for stellar absorption). Furthermore, metallicities based solely on the \niiha \ls local calibration could be severely overestimated, and these models allow us to determine the expected metallicities when shocks are present.

\begin{figure*}[t]
\centerline{
\includegraphics[width=5in]{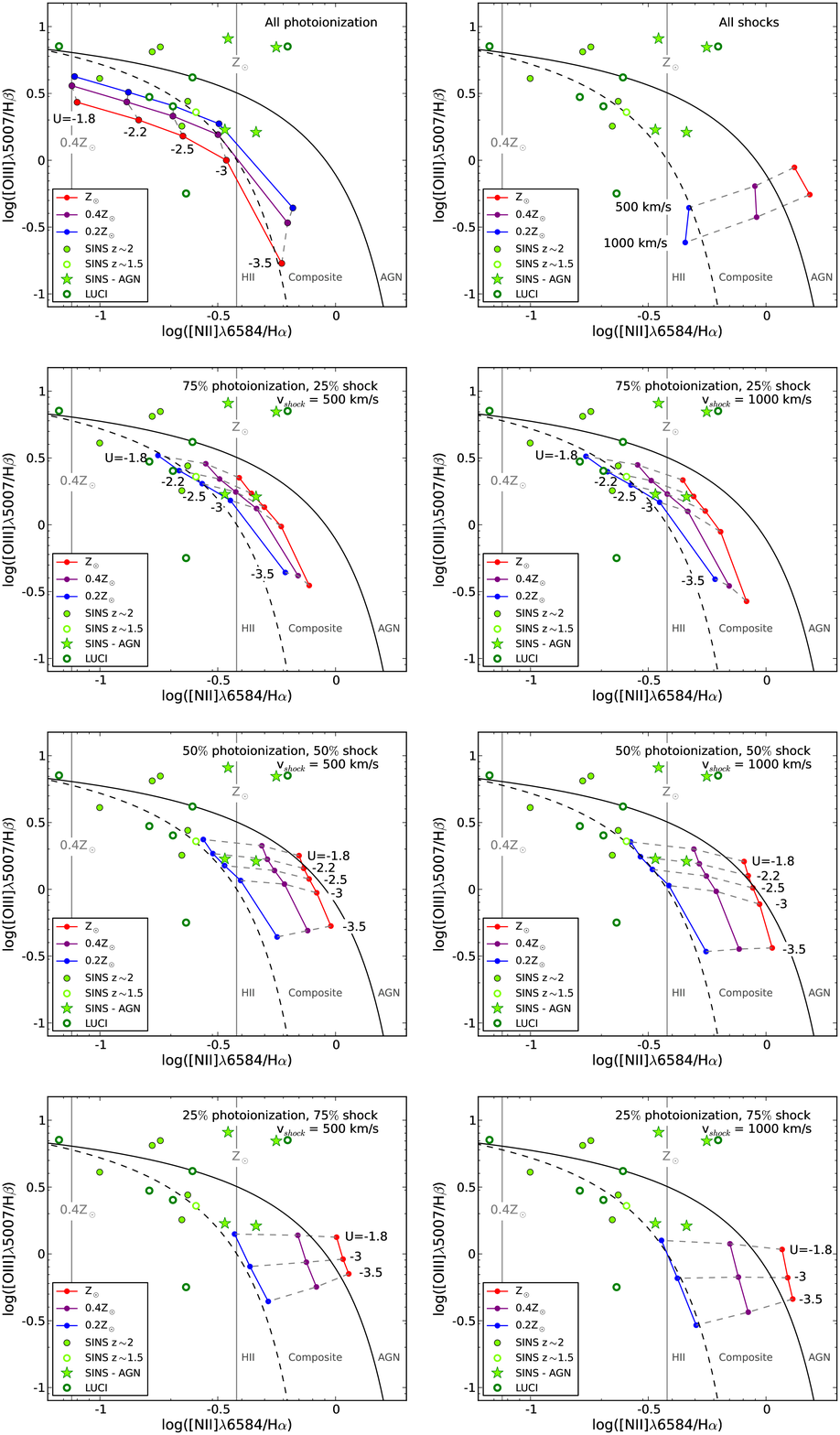}}
\caption{The BPT diagram with SINS and LUCI data as well as photoionization and shock models. The different panels represent different contribution fractions from photoionization and shocks, and different shock velocities. The colored lines represent different metallicities with red, purple and blue being Z$_{\odot}$, 0.4Z$_{\odot}$, and 0.2Z$_{\odot}$, respectively. The data points connected by grey dashed lines represent different ionization parameters in all but the upper right panel (where they represent different velocities), ranging from logU = -3.5 to -1.8. The vertical grey lines show the corresponding \niiha \ls values for Z$_{\odot}$ and 0.4Z$_{\odot}$ using the \cite{PetPag04} calibration. The combined photoionization and shock models occupy much of the `composite' region of the BPT diagram and thus these processes alone may be responsible for the positions of many \ztwo SFGs in the diagram, without the need to invoke an AGN.}
\end{figure*}

\subsection{Comparison of metallicity indicators}

A discussion of the mass-metallicity relation for the full SINS/zC-SINF and LUCI samples as well as resolved metallicity gradients will be discussed by \cite{Kur+13}. Here, we compare the gas phase metallicities derived using different nebular line ratios, all calibrated locally. \cite{KewEll08}, among others, have shown that the choice of metallicity calibration can have a large effect on the resulting metallicity, even for local SDSS galaxies. However, here we explore additional systematic discrepancies that can arise from differing ISM conditions in high-z galaxies.

In Figure 17, we have plotted the O3N2 versus N2-based metallicities of our galaxies using the \cite{PetPag04} calibrations. The deviation from the 1:1 relation is reminiscent of the offset above the SDSS star-forming sequence in the BPT diagram (Figure 6), and could reflect different physical conditions locally and at \ztwo or an obscured AGN contribution, driving the line excitation. Indeed, for AGN galaxies we find a mean offset between the gas-phase metallicities derived with the two different methods of 0.28dex (stddev = 0.12dex), while for the non-AGN galaxies, we find a mean offset of 0.16dex (stddev = 0.06dex). If we account for stellar \hb absorption, as previously shown in figure 9 for the BPT diagram, the deviation from the 1:1 relation decreases but the average offset is still 0.13dex (stddev = 0.06dex) for the non-AGN and 0.24 (stddev = 0.12dex) for the AGN galaxies, towards lower log(O/H) with the O3N2 calibration.

\begin{figure}[t]
\centerline{
\includegraphics[width=4in]{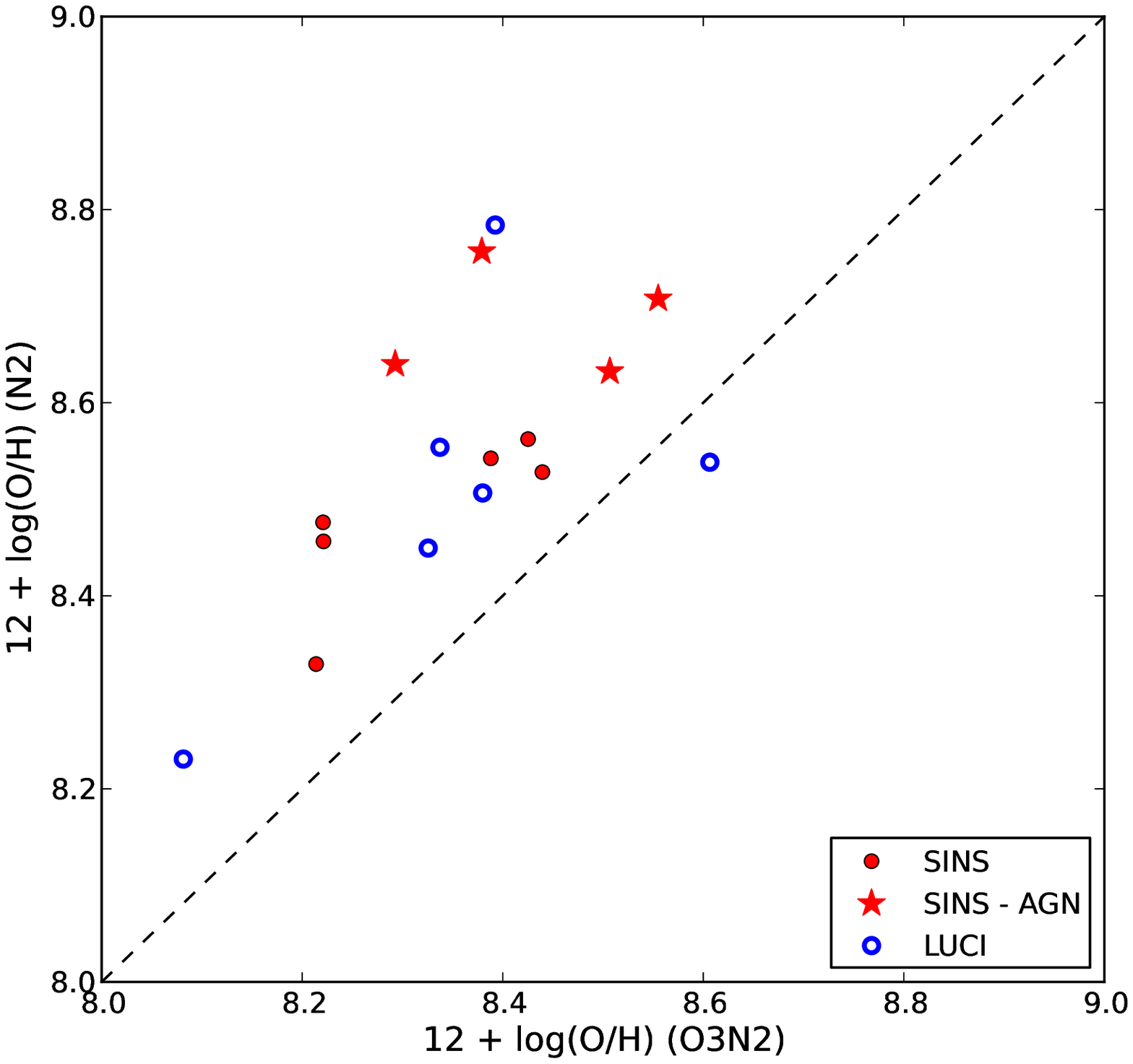}}
\caption{Comparison of gas phase metallicities calibrated using either the \niiha \ls calibration or the \oiiihb/\niiha \ls calibration \citep[both from][]{PetPag04}. The 1:1 relation is shown by the dashed line, SINS galaxies are shown as red symbols (filled circles for star-forming, and stars for AGN), and LUCI galaxies are shown as blue open circles. The N2 diagnostic is biased at \ztwo, systematically producing higher metallicities than the O3N2 diagnostic, even with the exclusion of AGN galaxies.}
\end{figure}

Thus, as shown by \cite{Liu+08}, some metallicity calibrations (e.g. N2) are more affected than others by the presence of shocks, AGN or higher interstellar pressure, and even while excluding the AGN galaxies, there is still an offset between the two calibrations. As mentioned in the previous section, using the O3N2 calibration brings the metallicity determinations into closer agreement with the photoionization and shock models presented here (as compared with the N2 calibration), but there is still some inconsistency. Despite the observed offsets, relative measurements among galaxies remain unaltered and the use of one or the other indicator would lead to similar conclusions.

Based on our findings here and in Section 3.2, it is possible that the mass-metallicity relation at \ztwo based on the \niiha \ls diagnostic could be overestimated by a factor of 2--3 in Z. This discrepancy may be even worse considering that we have excluded from this calculation AGN that have been identified solely based on spatially resolved observations.

\section{Conclusions}

Based on 22 z$\sim$1.4--2.5 SFGs from the SINS/zC-SINF and LUCI surveys (10 with spatially resolved data), we explore the effects of AGN, shocks and photoionization on line ratio diagnostics at \ztwo. In particular, our spatially-resolved data allow us to investigate how integrated line ratio measurements may be impacted by different excitation mechanisms. Our findings are as follows:
\begin{itemize}
\item With the use of spatially resolved line ratios, we are able to identify low-luminosity or obscured AGN that would otherwise not be detectable from integrated measurements, wherein the line ratios are the product of both a star-formation dominated disk and a highly excited nuclear region.
\item Shocks and photoionization may be responsible for the location of many \ztwo galaxies in the `composite' region of the BPT diagram. Shocks could contribute up to $\sim$25\% of the \ha \ls flux for our sample.
\item  In the composite region of the BPT diagram, there is a mix of galaxies with and without evidence for AGN and the prevalence of an AGN in this region appears to be correlated with \niiha (or similarly \mstar).
\item The MEx diagnostic \citep{Jun+11} does not accurately separate \ztwo star-forming and AGN galaxies according to the BPT diagram. However, applying a shift to \mstar \ls values based on the offset between the local and \ztwo mass-metallicity relations improves consistency of galaxy classification between the BPT and MEx diagrams.
\item Gas-phase metallicities that are determined from local calibrations appear biased to higher metallicities at high-z when the presence of photoionization and shocks are considered, by up to a factor of 3 (when using the N2 calibration) or 2 (when using the O3N2 calibration).
\end{itemize}

The key take-away messages from this paper are that one must be careful when using integrated line ratios, as there may be a large range in excitation that is hidden in the average spectrum and that the effects of shocks, photoionization and the presence of an AGN may all be responsible for elevated line ratios in \ztwo SFGs.

\acknowledgements
\small{We thank the ESO staff, especially those at Paranal Observatory, for their ongoing support during the many past and continuing observing runs over which the SINS project is being carried out as well as the LBTO staff. We also acknowledge the SINFONI, PARSEC and LUCI teams, whose devoted work on the instruments and lasers paved the way for the success of the SINS and LUCI observations. In addition, we thank Jonathan Trump for supplying the errors for his BPT data. SFN is supported by an NSF grfp grant. CM and AR and GZ acknowledge partial support by the ASI grant ``COFIS-Analisi Dati'' and by the INAF grant ``PRIN-2008'' and ``PRIN-2010''. }


\begin{thebibliography}{105}
\expandafter\ifx\csname natexlab\endcsname\relax\def\natexlab#1{#1}\fi

\bibitem[{{Abraham} {et~al.}(2004)}]{Abr+04}
{Abraham}, R.~G. {et~al.} 2004, \aj, 127, 2455

\bibitem[{{Adelberger} {et~al.}(2004){Adelberger}, {Steidel}, {Shapley},
  {Hunt}, {Erb}, {Reddy}, \& {Pettini}}]{Ade+04}
{Adelberger}, K.~L., {Steidel}, C.~C., {Shapley}, A.~E., {Hunt}, M.~P., {Erb},
  D.~K., {Reddy}, N.~A., \& {Pettini}, M. 2004, \apj, 607, 226

\bibitem[{{Ageorges} {et~al.}(2010){Ageorges}, {Seifert}, {J{\"u}tte},
  {Knierim}, {Lehmitz}, {Germeroth}, {Buschkamp}, {Polsterer}, {Pasquali},
  {Naranjo}, {Gemperlein}, {Hill}, {Feiz}, {Hofmann}, {Laun}, {Lederer},
  {Lenzen}, {Mall}, {Mandel}, {M{\"u}ller}, {Quirrenbach}, {Sch{\"a}ffner},
  {Storz}, \& {Weiser}}]{Age+10}
{Ageorges}, N., {Seifert}, W., {J{\"u}tte}, M., {Knierim}, V., {Lehmitz}, M.,
  {Germeroth}, A., {Buschkamp}, P., {Polsterer}, K., {Pasquali}, A., {Naranjo},
  V., {Gemperlein}, H., {Hill}, J., {Feiz}, C., {Hofmann}, R., {Laun}, W.,
  {Lederer}, R., {Lenzen}, R., {Mall}, U., {Mandel}, H., {M{\"u}ller}, P.,
  {Quirrenbach}, A., {Sch{\"a}ffner}, L., {Storz}, C., \& {Weiser}, P. 2010, in
  Society of Photo-Optical Instrumentation Engineers (SPIE) Conference Series,
  Vol. 7735, Society of Photo-Optical Instrumentation Engineers (SPIE)
  Conference Series

\bibitem[{{Allen} {et~al.}(2008){Allen}, {Groves}, {Dopita}, {Sutherland}, \&
  {Kewley}}]{All+08}
{Allen}, M.~G., {Groves}, B.~A., {Dopita}, M.~A., {Sutherland}, R.~S., \&
  {Kewley}, L.~J. 2008, \apjs, 178, 20

\bibitem[{{Asplund} {et~al.}(2009){Asplund}, {Grevesse}, {Sauval}, \&
  {Scott}}]{Asp+09}
{Asplund}, M., {Grevesse}, N., {Sauval}, A.~J., \& {Scott}, P. 2009, \araa, 47,
  481

\bibitem[{{Baldwin} {et~al.}(1981){Baldwin}, {Phillips}, \&
  {Terlevich}}]{BPT81}
{Baldwin}, J.~A., {Phillips}, M.~M., \& {Terlevich}, R. 1981, \pasp, 93, 5

\bibitem[{{Barger} {et~al.}(2008){Barger}, {Cowie}, \& {Wang}}]{Bar+08}
{Barger}, A.~J., {Cowie}, L.~L., \& {Wang}, W.-H. 2008, \apj, 689, 687

\bibitem[{{Berta} {et~al.}(2011){Berta}, {Magnelli}, {Nordon}, {Lutz}, {Wuyts},
  {Altieri}, {Andreani}, {Aussel}, {Casta{\~n}eda}, {Cepa}, {Cimatti}, {Daddi},
  {Elbaz}, {F{\"o}rster Schreiber}, {Genzel}, {Le Floc'h}, {Maiolino},
  {P{\'e}rez-Fournon}, {Poglitsch}, {Popesso}, {Pozzi}, {Riguccini},
  {Rodighiero}, {Sanchez-Portal}, {Sturm}, {Tacconi}, \& {Valtchanov}}]{Ber+11}
{Berta}, S., {Magnelli}, B., {Nordon}, R., {Lutz}, D., {Wuyts}, S., {Altieri},
  B., {Andreani}, P., {Aussel}, H., {Casta{\~n}eda}, H., {Cepa}, J., {Cimatti},
  A., {Daddi}, E., {Elbaz}, D., {F{\"o}rster Schreiber}, N.~M., {Genzel}, R.,
  {Le Floc'h}, E., {Maiolino}, R., {P{\'e}rez-Fournon}, I., {Poglitsch}, A.,
  {Popesso}, P., {Pozzi}, F., {Riguccini}, L., {Rodighiero}, G.,
  {Sanchez-Portal}, M., {Sturm}, E., {Tacconi}, L.~J., \& {Valtchanov}, I.
  2011, \aap, 532, A49

\bibitem[{{Bonzini} {et~al.}(2012){Bonzini}, {Mainieri}, {Padovani},
  {Kellermann}, {Miller}, {Rosati}, {Tozzi}, {Vattakunnel}, {Balestra},
  {Brandt}, {Luo}, \& {Xue}}]{Bon+12}
{Bonzini}, M., {Mainieri}, V., {Padovani}, P., {Kellermann}, K.~I., {Miller},
  N., {Rosati}, P., {Tozzi}, P., {Vattakunnel}, S., {Balestra}, I., {Brandt},
  W.~N., {Luo}, B., \& {Xue}, Y.~Q. 2012, \apjs, 203, 15

\bibitem[{{Brinchmann} {et~al.}(2008){Brinchmann}, {Pettini}, \&
  {Charlot}}]{Bri+08}
{Brinchmann}, J., {Pettini}, M., \& {Charlot}, S. 2008, \mnras, 385, 769

\bibitem[{{Bruzual} \& {Charlot}(2003)}]{BruCha03}
{Bruzual}, G. \& {Charlot}, S. 2003, \mnras, 344, 1000

\bibitem[{{Buschkamp} {et~al.}(2010){Buschkamp}, {Hofmann}, {Gemperlein},
  {Polsterer}, {Ageorges}, {Eisenhauer}, {Lederer}, {Honsberg}, {Haug}, {Eibl},
  {Seifert}, \& {Genzel}}]{Bus+10}
{Buschkamp}, P., {Hofmann}, R., {Gemperlein}, H., {Polsterer}, K., {Ageorges},
  N., {Eisenhauer}, F., {Lederer}, R., {Honsberg}, M., {Haug}, M., {Eibl}, J.,
  {Seifert}, W., \& {Genzel}, R. 2010, in Society of Photo-Optical
  Instrumentation Engineers (SPIE) Conference Series, Vol. 7735, Society of
  Photo-Optical Instrumentation Engineers (SPIE) Conference Series

\bibitem[{{Chabrier}(2003)}]{Cha03}
{Chabrier}, G. 2003, \pasp, 115, 763

\bibitem[{{Cimatti} {et~al.}(2002)}]{Cim+02}
{Cimatti}, A. {et~al.} 2002, \aap, 381, L68

\bibitem[{{Cowie} {et~al.}(1995){Cowie}, {Hu}, \& {Songaila}}]{CowHuSon95}
{Cowie}, L.~L., {Hu}, E.~M., \& {Songaila}, A. 1995, \aj, 110, 1576

\bibitem[{{Cresci} {et~al.}(2009)}]{Cre+09}
{Cresci}, G. {et~al.} 2009, \apj, 697, 115

\bibitem[{{Daddi} {et~al.}(2004)}]{Dad+04}
{Daddi}, E. {et~al.} 2004, \apjl, 600, L127

\bibitem[{{Daddi} {et~al.}(2007)}]{Dad+07}
---. 2007, \apj, 670, 156

\bibitem[{{Davies} {et~al.}(2011){Davies}, {F{\"o}rster Schreiber}, {Cresci},
  {Genzel}, {Bouch{\'e}}, {Burkert}, {Buschkamp}, {Genel}, {Hicks}, {Kurk},
  {Lutz}, {Newman}, {Shapiro}, {Sternberg}, {Tacconi}, \& {Wuyts}}]{Dav+11}
{Davies}, R., {F{\"o}rster Schreiber}, N.~M., {Cresci}, G., {Genzel}, R.,
  {Bouch{\'e}}, N., {Burkert}, A., {Buschkamp}, P., {Genel}, S., {Hicks}, E.,
  {Kurk}, J., {Lutz}, D., {Newman}, S., {Shapiro}, K., {Sternberg}, A.,
  {Tacconi}, L.~J., \& {Wuyts}, S. 2011, \apj, 741, 69

\bibitem[{{Davies}(2007)}]{Dav+07}
{Davies}, R.~I. 2007, \mnras, 375, 1099

\bibitem[{{Donley} {et~al.}(2012){Donley}, {Koekemoer}, {Brusa}, {Capak},
  {Cardamone}, {Civano}, {Ilbert}, {Impey}, {Kartaltepe}, {Miyaji}, {Salvato},
  {Sanders}, {Trump}, \& {Zamorani}}]{Don+12}
{Donley}, J.~L., {Koekemoer}, A.~M., {Brusa}, M., {Capak}, P., {Cardamone},
  C.~N., {Civano}, F., {Ilbert}, O., {Impey}, C.~D., {Kartaltepe}, J.~S.,
  {Miyaji}, T., {Salvato}, M., {Sanders}, D.~B., {Trump}, J.~R., \& {Zamorani},
  G. 2012, \apj, 748, 142

\bibitem[{{Dopita} {et~al.}(2006){Dopita}, {Fischera}, {Sutherland}, {Kewley},
  {Leitherer}, {Tuffs}, {Popescu}, {van Breugel}, \& {Groves}}]{Dop+06}
{Dopita}, M.~A., {Fischera}, J., {Sutherland}, R.~S., {Kewley}, L.~J.,
  {Leitherer}, C., {Tuffs}, R.~J., {Popescu}, C.~C., {van Breugel}, W., \&
  {Groves}, B.~A. 2006, \apjs, 167, 177

\bibitem[{{Dopita} {et~al.}(2000){Dopita}, {Kewley}, {Heisler}, \&
  {Sutherland}}]{Dop+00}
{Dopita}, M.~A., {Kewley}, L.~J., {Heisler}, C.~A., \& {Sutherland}, R.~S.
  2000, \apj, 542, 224

\bibitem[{{Dopita} \& {Sutherland}(1995)}]{DopSut95}
{Dopita}, M.~A. \& {Sutherland}, R.~S. 1995, \apj, 455, 468

\bibitem[{{Dufour} {et~al.}(1980){Dufour}, {Talbot}, {Jensen}, \&
  {Shields}}]{Duf+80}
{Dufour}, R.~J., {Talbot}, Jr., R.~J., {Jensen}, E.~B., \& {Shields}, G.~A.
  1980, \apj, 236, 119

\bibitem[{{Elmegreen} \& {Elmegreen}(2005)}]{ElmElm05}
{Elmegreen}, B.~G. \& {Elmegreen}, D.~M. 2005, \apj, 627, 632

\bibitem[{{Elmegreen} \& {Elmegreen}(2006)}]{ElmElm06}
---. 2006, \apj, 650, 644

\bibitem[{{Elmegreen} {et~al.}(2009){Elmegreen}, {Elmegreen}, {Marcus},
  {Shahinyan}, {Yau}, \& {Petersen}}]{Elm+09}
{Elmegreen}, D.~M., {Elmegreen}, B.~G., {Marcus}, M.~T., {Shahinyan}, K.,
  {Yau}, A., \& {Petersen}, M. 2009, \apj, 701, 306

\bibitem[{{Elmegreen} {et~al.}(2004){Elmegreen}, {Elmegreen}, \&
  {Sheets}}]{ElmElmShe04}
{Elmegreen}, D.~M., {Elmegreen}, B.~G., \& {Sheets}, C.~M. 2004, \apj, 603, 74

\bibitem[{{Epinat} {et~al.}(2009){Epinat}, {Contini}, {Le F{\`e}vre},
  {Vergani}, {Garilli}, {Amram}, {Queyrel}, {Tasca}, \& {Tresse}}]{Epi+09}
{Epinat}, B., {Contini}, T., {Le F{\`e}vre}, O., {Vergani}, D., {Garilli}, B.,
  {Amram}, P., {Queyrel}, J., {Tasca}, L., \& {Tresse}, L. 2009, \aap, 504, 789

\bibitem[{{Epinat} {et~al.}(2012){Epinat}, {Tasca}, {Amram}, {Contini}, {Le
  F{\`e}vre}, {Queyrel}, {Vergani}, {Garilli}, {Kissler-Patig}, {Moultaka},
  {Paioro}, {Tresse}, {Bournaud}, {L{\'o}pez-Sanjuan}, \& {Perret}}]{Epi+12}
{Epinat}, B., {Tasca}, L., {Amram}, P., {Contini}, T., {Le F{\`e}vre}, O.,
  {Queyrel}, J., {Vergani}, D., {Garilli}, B., {Kissler-Patig}, M., {Moultaka},
  J., {Paioro}, L., {Tresse}, L., {Bournaud}, F., {L{\'o}pez-Sanjuan}, C., \&
  {Perret}, V. 2012, \aap, 539, A92

\bibitem[{{Erb} {et~al.}(2006{\natexlab{a}}){Erb}, {Shapley}, {Pettini},
  {Steidel}, {Reddy}, \& {Adelberger}}]{Erb+06a}
{Erb}, D.~K., {Shapley}, A.~E., {Pettini}, M., {Steidel}, C.~C., {Reddy},
  N.~A., \& {Adelberger}, K.~L. 2006{\natexlab{a}}, \apj, 644, 813

\bibitem[{{Erb} {et~al.}(2003){Erb}, {Shapley}, {Steidel}, {Pettini},
  {Adelberger}, {Hunt}, {Moorwood}, \& {Cuby}}]{Erb+03}
{Erb}, D.~K., {Shapley}, A.~E., {Steidel}, C.~C., {Pettini}, M., {Adelberger},
  K.~L., {Hunt}, M.~P., {Moorwood}, A.~F.~M., \& {Cuby}, J.-G. 2003, \apj, 591,
  101

\bibitem[{{Erb} {et~al.}(2006{\natexlab{b}}){Erb}, {Steidel}, {Shapley},
  {Pettini}, {Reddy}, \& {Adelberger}}]{Erb+06}
{Erb}, D.~K., {Steidel}, C.~C., {Shapley}, A.~E., {Pettini}, M., {Reddy},
  N.~A., \& {Adelberger}, K.~L. 2006{\natexlab{b}}, \apj, 646, 107

\bibitem[{{Ferland} {et~al.}(2013){Ferland}, {Porter}, {van Hoof}, {Williams},
  {Abel}, {Lykins}, {Shaw}, {Henney}, \& {Stancil}}]{Fer+13}
{Ferland}, G.~J., {Porter}, R.~L., {van Hoof}, P.~A.~M., {Williams}, R.~J.~R.,
  {Abel}, N.~P., {Lykins}, M.~L., {Shaw}, G., {Henney}, W.~J., \& {Stancil},
  P.~C. 2013, RMxAA, 49, 137

\bibitem[{{F{\"o}rster Schreiber} {et~al.}(2011{\natexlab{a}}){F{\"o}rster
  Schreiber}, {Shapley}, {Erb}, {Genzel}, {Steidel}, {Bouch{\'e}}, {Cresci}, \&
  {Davies}}]{For+11a}
{F{\"o}rster Schreiber}, N.~M., {Shapley}, A.~E., {Erb}, D.~K., {Genzel}, R.,
  {Steidel}, C.~C., {Bouch{\'e}}, N., {Cresci}, G., \& {Davies}, R.
  2011{\natexlab{a}}, \apj, 731, 65

\bibitem[{{F{\"o}rster Schreiber} {et~al.}(2011{\natexlab{b}}){F{\"o}rster
  Schreiber}, {Shapley}, {Genzel}, {Bouch{\'e}}, {Cresci}, {Davies}, {Erb},
  {Genel}, {Lutz}, {Newman}, {Shapiro}, {Steidel}, {Sternberg}, \&
  {Tacconi}}]{For+11b}
{F{\"o}rster Schreiber}, N.~M., {Shapley}, A.~E., {Genzel}, R., {Bouch{\'e}},
  N., {Cresci}, G., {Davies}, R., {Erb}, D.~K., {Genel}, S., {Lutz}, D.,
  {Newman}, S., {Shapiro}, K.~L., {Steidel}, C.~C., {Sternberg}, A., \&
  {Tacconi}, L.~J. 2011{\natexlab{b}}, \apj, 739, 45

\bibitem[{{F{\"o}rster Schreiber} {et~al.}(2006)}]{For+06}
{F{\"o}rster Schreiber}, N.~M. {et~al.} 2006, \apj, 645, 1062

\bibitem[{{F{\"o}rster Schreiber} {et~al.}(2009)}]{For+09}
---. 2009, \apj, 706, 1364

\bibitem[{{F{\"o}rster Schreiber} {et~al.}(2013a)}]{For+13a}
{F{\"o}rster Schreiber}, N.~M. {et~al.} 2013a, in preparation

\bibitem[{{F{\"o}rster Schreiber} {et~al.}(2013b)}]{For+13b}
{F{\"o}rster Schreiber}, N.~M. {et~al.} 2013b, in preparation

\bibitem[{{Genzel} {et~al.}(2011){Genzel}, {Newman}, {Jones}, {F{\"o}rster
  Schreiber}, {Shapiro}, {Genel}, {Lilly}, {Renzini}, {Tacconi}, {Bouch{\'e}},
  {Burkert}, {Cresci}, {Buschkamp}, {Carollo}, {Ceverino}, {Davies}, {Dekel},
  {Eisenhauer}, {Hicks}, {Kurk}, {Lutz}, {Mancini}, {Naab}, {Peng},
  {Sternberg}, {Vergani}, \& {Zamorani}}]{Gen+11}
{Genzel}, R., {Newman}, S., {Jones}, T., {F{\"o}rster Schreiber}, N.~M.,
  {Shapiro}, K., {Genel}, S., {Lilly}, S.~J., {Renzini}, A., {Tacconi}, L.~J.,
  {Bouch{\'e}}, N., {Burkert}, A., {Cresci}, G., {Buschkamp}, P., {Carollo},
  C.~M., {Ceverino}, D., {Davies}, R., {Dekel}, A., {Eisenhauer}, F., {Hicks},
  E., {Kurk}, J., {Lutz}, D., {Mancini}, C., {Naab}, T., {Peng}, Y.,
  {Sternberg}, A., {Vergani}, D., \& {Zamorani}, G. 2011, \apj, 733, 101

\bibitem[{{Genzel} {et~al.}(2006)}]{Gen+06}
{Genzel}, R. {et~al.} 2006, \nat, 442, 786

\bibitem[{{Genzel} {et~al.}(2008)}]{Gen+08}
---. 2008, \apj, 687, 59

\bibitem[{{Gnat} \& {Sternberg}(2009)}]{GnaSte09}
{Gnat}, O. \& {Sternberg}, A. 2009, \apj, 693, 1514

\bibitem[{{Hainline} {et~al.}(2012){Hainline}, {Shapley}, {Greene}, {Steidel},
  {Reddy}, \& {Erb}}]{Hai+12}
{Hainline}, K.~N., {Shapley}, A.~E., {Greene}, J.~E., {Steidel}, C.~C.,
  {Reddy}, N.~A., \& {Erb}, D.~K. 2012, \apj, 760, 74

\bibitem[{{Hayashi} {et~al.}(2009){Hayashi}, {Motohara}, {Shimasaku},
  {Onodera}, {Uchimoto}, {Kashikawa}, {Yoshida}, {Okamura}, {Ly}, \&
  {Malkan}}]{Hay+09}
{Hayashi}, M., {Motohara}, K., {Shimasaku}, K., {Onodera}, M., {Uchimoto},
  Y.~K., {Kashikawa}, N., {Yoshida}, M., {Okamura}, S., {Ly}, C., \& {Malkan},
  M.~A. 2009, \apj, 691, 140

\bibitem[{{Hillier} \& {Miller}(1998)}]{HilMil98}
{Hillier}, D.~J. \& {Miller}, D.~L. 1998, \apj, 496, 407

\bibitem[{{Jones} {et~al.}(2010){Jones}, {Ellis}, {Jullo}, \&
  {Richard}}]{Jon+10}
{Jones}, T., {Ellis}, R., {Jullo}, E., \& {Richard}, J. 2010, \apjl, 725, L176

\bibitem[{{Jones} {et~al.}(2012){Jones}, {Stark}, \& {Ellis}}]{Jon+12}
{Jones}, T., {Stark}, D.~P., \& {Ellis}, R.~S. 2012, \apj, 751, 51

\bibitem[{{Juneau} {et~al.}(2011){Juneau}, {Dickinson}, {Alexander}, \&
  {Salim}}]{Jun+11}
{Juneau}, S., {Dickinson}, M., {Alexander}, D.~M., \& {Salim}, S. 2011, \apj,
  736, 104

\bibitem[{{Juneau} {et~al.}(2013){Juneau}, {Dickinson}, {Bournaud},
  {Alexander}, {Daddi}, {Mullaney}, {Magnelli}, {Kartaltepe}, {Hwang},
  {Willner}, {Coil}, {Rosario}, {Trump}, {Weiner}, {Willmer}, {Cooper},
  {Elbaz}, {Faber}, {Frayer}, {Kocevski}, {Laird}, {Monkiewicz}, {Nandra},
  {Newman}, {Salim}, \& {Symeonidis}}]{Jun+13}
{Juneau}, S., {Dickinson}, M., {Bournaud}, F., {Alexander}, D.~M., {Daddi}, E.,
  {Mullaney}, J.~R., {Magnelli}, B., {Kartaltepe}, J.~S., {Hwang}, H.~S.,
  {Willner}, S.~P., {Coil}, A.~L., {Rosario}, D.~J., {Trump}, J.~R., {Weiner},
  B.~J., {Willmer}, C.~N.~A., {Cooper}, M.~C., {Elbaz}, D., {Faber}, S.~M.,
  {Frayer}, D.~T., {Kocevski}, D.~D., {Laird}, E.~S., {Monkiewicz}, J.~A.,
  {Nandra}, K., {Newman}, J.~A., {Salim}, S., \& {Symeonidis}, M. 2013, \apj,
  764, 176

\bibitem[{{Kauffmann} {et~al.}(2003){Kauffmann}, {Heckman}, {Tremonti},
  {Brinchmann}, {Charlot}, {White}, {Ridgway}, {Brinkmann}, {Fukugita}, {Hall},
  {Ivezi{\'c}}, {Richards}, \& {Schneider}}]{Kau+03}
{Kauffmann}, G., {Heckman}, T.~M., {Tremonti}, C., {Brinchmann}, J., {Charlot},
  S., {White}, S.~D.~M., {Ridgway}, S.~E., {Brinkmann}, J., {Fukugita}, M.,
  {Hall}, P.~B., {Ivezi{\'c}}, {\v Z}., {Richards}, G.~T., \& {Schneider},
  D.~P. 2003, \mnras, 346, 1055

\bibitem[{{Kewley} {et~al.}(2001){Kewley}, {Dopita},
  {Sutherland}, {Heisler}, \& {Trevena}}]{Kew+01a}
{Kewley}, L.~J., {Dopita}, M.~A., {Sutherland}, R.~S., {Heisler}, C.~A., \&
  {Trevena}, J. 2001, \apj, 556, 121

\bibitem[{{Kewley} \& {Ellison}(2008)}]{KewEll08}
{Kewley}, L.~J. \& {Ellison}, S.~L. 2008, \apj, 681, 1183

\bibitem[{{Kewley} {et~al.}(2006){Kewley}, {Groves}, {Kauffmann}, \&
  {Heckman}}]{Kew+06}
{Kewley}, L.~J., {Groves}, B., {Kauffmann}, G., \& {Heckman}, T. 2006, \mnras,
  372, 961

\bibitem[{{Komatsu} {et~al.}(2011){Komatsu}, {Smith}, {Dunkley}, {Bennett},
  {Gold}, {Hinshaw}, {Jarosik}, {Larson}, {Nolta}, {Page}, {Spergel},
  {Halpern}, {Hill}, {Kogut}, {Limon}, {Meyer}, {Odegard}, {Tucker}, {Weiland},
  {Wollack}, \& {Wright}}]{Kom+11}
{Komatsu}, E., {Smith}, K.~M., {Dunkley}, J., {Bennett}, C.~L., {Gold}, B.,
  {Hinshaw}, G., {Jarosik}, N., {Larson}, D., {Nolta}, M.~R., {Page}, L.,
  {Spergel}, D.~N., {Halpern}, M., {Hill}, R.~S., {Kogut}, A., {Limon}, M.,
  {Meyer}, S.~S., {Odegard}, N., {Tucker}, G.~S., {Weiland}, J.~L., {Wollack},
  E., \& {Wright}, E.~L. 2011, \apjs, 192, 18

\bibitem[{{Kong} {et~al.}(2006)}]{Kon+06}
{Kong}, X. {et~al.} 2006, \apj, 638, 72

\bibitem[{{Kriek} {et~al.}(2009){Kriek}, {van Dokkum}, {Labb{\'e}}, {Franx},
  {Illingworth}, {Marchesini}, \& {Quadri}}]{Kri+09}
{Kriek}, M., {van Dokkum}, P.~G., {Labb{\'e}}, I., {Franx}, M., {Illingworth},
  G.~D., {Marchesini}, D., \& {Quadri}, R.~F. 2009, \apj, 700, 221

\bibitem[{{Kriek} {et~al.}(2007)}]{Kri+07}
{Kriek}, M. {et~al.} 2007, \apj, 669, 776

\bibitem[{{Kroupa}(2001)}]{Kro01}
{Kroupa}, P. 2001, \mnras, 322, 231

\bibitem[{{Kurk} {et~al.}(2013)}]{Kur+13}
{Kurk}, J. {et~al.} 2013, in preparation

\bibitem[{{Lacy} {et~al.}(2007){Lacy}, {Petric}, {Sajina}, {Canalizo},
  {Storrie-Lombardi}, {Armus}, {Fadda}, \& {Marleau}}]{Lac+07}
{Lacy}, M., {Petric}, A.~O., {Sajina}, A., {Canalizo}, G., {Storrie-Lombardi},
  L.~J., {Armus}, L., {Fadda}, D., \& {Marleau}, F.~R. 2007, \aj, 133, 186

\bibitem[{{Law} {et~al.}(2007){Law}, {Steidel}, {Erb}, {Larkin}, {Pettini},
  {Shapley}, \& {Wright}}]{Law+07}
{Law}, D.~R., {Steidel}, C.~C., {Erb}, D.~K., {Larkin}, J.~E., {Pettini}, M.,
  {Shapley}, A.~E., \& {Wright}, S.~A. 2007, \apj, 669, 929

\bibitem[{{Law} {et~al.}(2009){Law}, {Steidel}, {Erb}, {Larkin}, {Pettini},
  {Shapley}, \& {Wright}}]{Law+09}
---. 2009, \apj, 697, 2057

\bibitem[{{Law} {et~al.}(2012){Law}, {Steidel}, {Shapley}, {Nagy}, {Reddy}, \&
  {Erb}}]{Law+12}
{Law}, D.~R., {Steidel}, C.~C., {Shapley}, A.~E., {Nagy}, S.~R., {Reddy},
  N.~A., \& {Erb}, D.~K. 2012, ArXiv e-prints

\bibitem[{{Leitherer} {et~al.}(1999){Leitherer}, {Schaerer}, {Goldader},
  {Gonz{\'a}lez Delgado}, {Robert}, {Kune}, {de Mello}, {Devost}, \&
  {Heckman}}]{Lei+99}
{Leitherer}, C., {Schaerer}, D., {Goldader}, J.~D., {Gonz{\'a}lez Delgado},
  R.~M., {Robert}, C., {Kune}, D.~F., {de Mello}, D.~F., {Devost}, D., \&
  {Heckman}, T.~M. 1999, \apjs, 123, 3

\bibitem[{{Lemoine-Busserolle} \& {Lamareille}(2010)}]{LemLam10}
{Lemoine-Busserolle}, M. \& {Lamareille}, F. 2010, \mnras, 402, 2291

\bibitem[{{Levesque} {et~al.}(2010){Levesque}, {Kewley}, \& {Larson}}]{Lev+10}
{Levesque}, E.~M., {Kewley}, L.~J., \& {Larson}, K.~L. 2010, \aj, 139, 712

\bibitem[{{Lilly} {et~al.}(2013){Lilly}, {Carollo}, {Pipino}, {Renzini}, \&
  {Peng}}]{Lil+13}
{Lilly}, S.~J., {Carollo}, C.~M., {Pipino}, A., {Renzini}, A., \& {Peng}, Y.
  2013, ArXiv e-prints

\bibitem[{{Lilly} {et~al.}(2007)}]{Lil+07}
{Lilly}, S.~J. {et~al.} 2007, \apjs, 172, 70

\bibitem[{{Liu} {et~al.}(2008){Liu}, {Shapley}, {Coil}, {Brinchmann}, \&
  {Ma}}]{Liu+08}
{Liu}, X., {Shapley}, A.~E., {Coil}, A.~L., {Brinchmann}, J., \& {Ma}, C.-P.
  2008, \apj, 678, 758

\bibitem[{{Mancini} {et~al.}(2011){Mancini}, {Foerster Schreiber}, {Renzini},
  {Cresci}, {Hicks}, {Peng}, {Vergani}, {Lilly}, {Carollo}, {Pozzetti},
  {Zamorani}, {Daddi}, {Genzel}, {Maraston}, {McCracken}, {Tacconi}, {Bouche},
  {Davies}, {Oesch}, {Shapiro}, {Mainieri}, {Lutz}, {Mignoli}, \&
  {Sternberg}}]{Man+11}
{Mancini}, C., {Foerster Schreiber}, N., {Renzini}, A., {Cresci}, G., {Hicks},
  E., {Peng}, Y., {Vergani}, D., {Lilly}, S., {Carollo}, C.~M., {Pozzetti}, L.,
  {Zamorani}, G., {Daddi}, E., {Genzel}, R., {Maraston}, C., {McCracken},
  H.~J., {Tacconi}, L.~J., {Bouche}, N., {Davies}, R.~I., {Oesch}, P.,
  {Shapiro}, K., {Mainieri}, V., {Lutz}, D., {Mignoli}, M., \& {Sternberg}, A.
  2011, ArXiv e-prints

\bibitem[{{Meynet} {et~al.}(1994){Meynet}, {Maeder}, {Schaller}, {Schaerer}, \&
  {Charbonnel}}]{Mey+94}
{Meynet}, G., {Maeder}, A., {Schaller}, G., {Schaerer}, D., \& {Charbonnel}, C.
  1994, \aaps, 103, 97

\bibitem[{{Mignoli} {et~al.}(2005){Mignoli}, {Cimatti}, {Zamorani}, {Pozzetti},
  {Daddi}, {Renzini}, {Broadhurst}, {Cristiani}, {D'Odorico}, {Fontana},
  {Giallongo}, {Gilmozzi}, {Menci}, \& {Saracco}}]{Mig+05}
{Mignoli}, M., {Cimatti}, A., {Zamorani}, G., {Pozzetti}, L., {Daddi}, E.,
  {Renzini}, A., {Broadhurst}, T., {Cristiani}, S., {D'Odorico}, S., {Fontana},
  A., {Giallongo}, E., {Gilmozzi}, R., {Menci}, N., \& {Saracco}, P. 2005,
  \aap, 437, 883

\bibitem[{{Miller} {et~al.}(2008){Miller}, {Fomalont}, {Kellermann},
  {Mainieri}, {Norman}, {Padovani}, {Rosati}, \& {Tozzi}}]{Mil+08}
{Miller}, N.~A., {Fomalont}, E.~B., {Kellermann}, K.~I., {Mainieri}, V.,
  {Norman}, C., {Padovani}, P., {Rosati}, P., \& {Tozzi}, P. 2008, \apjs, 179,
  114

\bibitem[{{Newman} {et~al.}(2013){Newman}, {Genzel}, {F{\"o}rster Schreiber},
  {Shapiro Griffin}, {Mancini}, {Lilly}, {Renzini}, {Bouch{\'e}}, {Burkert},
  {Buschkamp}, {Carollo}, {Cresci}, {Davies}, {Eisenhauer}, {Genel}, {Hicks},
  {Kurk}, {Lutz}, {Naab}, {Peng}, {Sternberg}, {Tacconi}, {Wuyts}, {Zamorani},
  \& {Vergani}}]{New+13a}
{Newman}, S.~F., {Genzel}, R., {F{\"o}rster Schreiber}, N.~M., {Shapiro
  Griffin}, K., {Mancini}, C., {Lilly}, S.~J., {Renzini}, A., {Bouch{\'e}}, N.,
  {Burkert}, A., {Buschkamp}, P., {Carollo}, C.~M., {Cresci}, G., {Davies}, R.,
  {Eisenhauer}, F., {Genel}, S., {Hicks}, E.~K.~S., {Kurk}, J., {Lutz}, D.,
  {Naab}, T., {Peng}, Y., {Sternberg}, A., {Tacconi}, L.~J., {Wuyts}, S.,
  {Zamorani}, G., \& {Vergani}, D. 2013, \apj, 767, 104

\bibitem[{{Newman} {et~al.}(2012){Newman}, {Shapiro Griffin}, {Genzel},
  {Davies}, {F{\"o}rster-Schreiber}, {Tacconi}, {Kurk}, {Wuyts}, {Genel},
  {Lilly}, {Renzini}, {Bouch{\'e}}, {Burkert}, {Cresci}, {Buschkamp},
  {Carollo}, {Eisenhauer}, {Hicks}, {Lutz}, {Mancini}, {Naab}, {Peng}, \&
  {Vergani}}]{New+12a}
{Newman}, S.~F., {Shapiro Griffin}, K., {Genzel}, R., {Davies}, R.,
  {F{\"o}rster-Schreiber}, N.~M., {Tacconi}, L.~J., {Kurk}, J., {Wuyts}, S.,
  {Genel}, S., {Lilly}, S.~J., {Renzini}, A., {Bouch{\'e}}, N., {Burkert}, A.,
  {Cresci}, G., {Buschkamp}, P., {Carollo}, C.~M., {Eisenhauer}, F., {Hicks},
  E., {Lutz}, D., {Mancini}, C., {Naab}, T., {Peng}, Y., \& {Vergani}, D. 2012,
  \apj, 752, 111

\bibitem[{{Pauldrach} {et~al.}(2001){Pauldrach}, {Hoffmann}, \&
  {Lennon}}]{Pau+01}
{Pauldrach}, A.~W.~A., {Hoffmann}, T.~L., \& {Lennon}, M. 2001, \aap, 375, 161

\bibitem[{{Peng} {et~al.}(2002){Peng}, {Ho}, {Impey}, \& {Rix}}]{Pen+02}
{Peng}, C.~Y., {Ho}, L.~C., {Impey}, C.~D., \& {Rix}, H.-W. 2002, \aj, 124, 266

\bibitem[{{Pettini} \& {Pagel}(2004)}]{PetPag04}
{Pettini}, M. \& {Pagel}, B.~E.~J. 2004, \mnras, 348, L59

\bibitem[{{Rangel} {et~al.}(2013){Rangel}, {Nandra}, {Laird}, \&
  {Orange}}]{Ran+13}
{Rangel}, C., {Nandra}, K., {Laird}, E.~S., \& {Orange}, P. 2013, \mnras, 428,
  3089

\bibitem[{{Renzini} \& {da Costa}(1997)}]{Ren+97}
{Renzini}, A. \& {da Costa}, L.~N. 1997, The Messenger, 87, 23

\bibitem[{{Rich} {et~al.}(2011){Rich}, {Kewley}, \& {Dopita}}]{Ric+11}
{Rich}, J.~A., {Kewley}, L.~J., \& {Dopita}, M.~A. 2011, \apj, 734, 87

\bibitem[{{Rodighiero} {et~al.}(2011){Rodighiero}, {Daddi}, {Baronchelli},
  {Cimatti}, {Renzini}, {Aussel}, {Popesso}, {Lutz}, {Andreani}, {Berta},
  {Cava}, {Elbaz}, {Feltre}, {Fontana}, {F{\"o}rster Schreiber},
  {Franceschini}, {Genzel}, {Grazian}, {Gruppioni}, {Ilbert}, {Le Floch},
  {Magdis}, {Magliocchetti}, {Magnelli}, {Maiolino}, {McCracken}, {Nordon},
  {Poglitsch}, {Santini}, {Pozzi}, {Riguccini}, {Tacconi}, {Wuyts}, \&
  {Zamorani}}]{Rod+11}
{Rodighiero}, G., {Daddi}, E., {Baronchelli}, I., {Cimatti}, A., {Renzini}, A.,
  {Aussel}, H., {Popesso}, P., {Lutz}, D., {Andreani}, P., {Berta}, S., {Cava},
  A., {Elbaz}, D., {Feltre}, A., {Fontana}, A., {F{\"o}rster Schreiber}, N.~M.,
  {Franceschini}, A., {Genzel}, R., {Grazian}, A., {Gruppioni}, C., {Ilbert},
  O., {Le Floch}, E., {Magdis}, G., {Magliocchetti}, M., {Magnelli}, B.,
  {Maiolino}, R., {McCracken}, H., {Nordon}, R., {Poglitsch}, A., {Santini},
  P., {Pozzi}, F., {Riguccini}, L., {Tacconi}, L.~J., {Wuyts}, S., \&
  {Zamorani}, G. 2011, \apjl, 739, L40+

\bibitem[{{Seifert} {et~al.}(2010){Seifert}, {Ageorges}, {Lehmitz},
  {Buschkamp}, {Knierim}, {Polsterer}, {Germeroth}, {Pasquali}, {Naranjo},
  {J{\"u}tte}, {Feiz}, {Gemperlein}, {Hofmann}, {Laun}, {Lederer}, {Lenzen},
  {Mall}, {Mandel}, {M{\"u}ller}, {Quirrenbach}, {Sch{\"a}ffner}, {Storz}, \&
  {Weiser}}]{Sei+10}
{Seifert}, W., {Ageorges}, N., {Lehmitz}, M., {Buschkamp}, P., {Knierim}, V.,
  {Polsterer}, K., {Germeroth}, A., {Pasquali}, A., {Naranjo}, V., {J{\"u}tte},
  M., {Feiz}, C., {Gemperlein}, H., {Hofmann}, R., {Laun}, W., {Lederer}, R.,
  {Lenzen}, R., {Mall}, U., {Mandel}, H., {M{\"u}ller}, P., {Quirrenbach}, A.,
  {Sch{\"a}ffner}, L., {Storz}, C., \& {Weiser}, P. 2010, in Society of
  Photo-Optical Instrumentation Engineers (SPIE) Conference Series, Vol. 7735,
  Society of Photo-Optical Instrumentation Engineers (SPIE) Conference Series

\bibitem[{{Shapiro} {et~al.}(2008)}]{Sha+08}
{Shapiro}, K.~L. {et~al.} 2008, \apj, 682, 231

\bibitem[{{Shapley} {et~al.}(2005){Shapley}, {Steidel}, {Erb}, {Reddy},
  {Adelberger}, {Pettini}, {Barmby}, \& {Huang}}]{Shapley+05}
{Shapley}, A.~E., {Steidel}, C.~C., {Erb}, D.~K., {Reddy}, N.~A., {Adelberger},
  K.~L., {Pettini}, M., {Barmby}, P., \& {Huang}, J. 2005, \apj, 626, 698

\bibitem[{{Steidel} {et~al.}(2004){Steidel}, {Shapley}, {Pettini},
  {Adelberger}, {Erb}, {Reddy}, \& {Hunt}}]{Ste+04}
{Steidel}, C.~C., {Shapley}, A.~E., {Pettini}, M., {Adelberger}, K.~L., {Erb},
  D.~K., {Reddy}, N.~A., \& {Hunt}, M.~P. 2004, \apj, 604, 534

\bibitem[{{Stern} {et~al.}(2005){Stern}, {Eisenhardt}, {Gorjian}, {Kochanek},
  {Caldwell}, {Eisenstein}, {Brodwin}, {Brown}, {Cool}, {Dey}, {Green},
  {Jannuzi}, {Murray}, {Pahre}, \& {Willner}}]{Stern+05}
{Stern}, D., {Eisenhardt}, P., {Gorjian}, V., {Kochanek}, C.~S., {Caldwell},
  N., {Eisenstein}, D., {Brodwin}, M., {Brown}, M.~J.~I., {Cool}, R., {Dey},
  A., {Green}, P., {Jannuzi}, B.~T., {Murray}, S.~S., {Pahre}, M.~A., \&
  {Willner}, S.~P. 2005, \apj, 631, 163

\bibitem[{{Swinbank} {et~al.}(2012{\natexlab{a}}){Swinbank}, {Sobral}, {Smail},
  {Geach}, {Best}, {McCarthy}, {Crain}, \& {Theuns}}]{Swi+12a}
{Swinbank}, A.~M., {Sobral}, D., {Smail}, I., {Geach}, J.~E., {Best}, P.~N.,
  {McCarthy}, I.~G., {Crain}, R.~A., \& {Theuns}, T. 2012{\natexlab{a}},
  \mnras, 426, 935

\bibitem[{{Swinbank} {et~al.}(2012{\natexlab{b}}){Swinbank}, {Smail}, {Sobral},
  {Theuns}, {Best}, \& {Geach}}]{Swi+12b}
{Swinbank}, M., {Smail}, I., {Sobral}, D., {Theuns}, T., {Best}, P., \&
  {Geach}, J. 2012{\natexlab{b}}, ArXiv e-prints

\bibitem[{{Tecza} {et~al.}(2004)}]{Tec+04}
{Tecza}, M. {et~al.} 2004, \apjl, 605, L109

\bibitem[{{Tremonti} {et~al.}(2004)}]{Tre+04}
{Tremonti}, C.~A. {et~al.} 2004, \apj, 613, 898

\bibitem[{{Trump} {et~al.}(2013){Trump}, {Konidaris}, {Barro}, {Koo},
  {Kocevski}, {Juneau}, {Weiner}, {Faber}, {McLean}, {Yan},
  {P{\'e}rez-Gonz{\'a}lez}, \& {Villar}}]{Tru+13}
{Trump}, J.~R., {Konidaris}, N.~P., {Barro}, G., {Koo}, D.~C., {Kocevski},
  D.~D., {Juneau}, S., {Weiner}, B.~J., {Faber}, S.~M., {McLean}, I.~S., {Yan},
  R., {P{\'e}rez-Gonz{\'a}lez}, P.~G., \& {Villar}, V. 2013, \apjl, 763, L6

\bibitem[{{van den Bergh} {et~al.}(1996){van den Bergh}, {Abraham}, {Ellis},
  {Tanvir}, {Santiago}, \& {Glazebrook}}]{vdBer+96}
{van den Bergh}, S., {Abraham}, R.~G., {Ellis}, R.~S., {Tanvir}, N.~R.,
  {Santiago}, B.~X., \& {Glazebrook}, K.~G. 1996, \aj, 112, 359

\bibitem[{{van Dokkum} {et~al.}(2005){van Dokkum}, {Kriek}, {Rodgers}, {Franx},
  \& {Puxley}}]{vDok+05}
{van Dokkum}, P.~G., {Kriek}, M., {Rodgers}, B., {Franx}, M., \& {Puxley}, P.
  2005, \apjl, 622, L13

\bibitem[{{van Starkenburg} {et~al.}(2008){van Starkenburg}, {van der Werf},
  {Franx}, {Labb{\'e}}, {Rudnick}, \& {Wuyts}}]{vSta+08}
{van Starkenburg}, L., {van der Werf}, P.~P., {Franx}, M., {Labb{\'e}}, I.,
  {Rudnick}, G., \& {Wuyts}, S. 2008, \aap, 488, 99

\bibitem[{{V{\'a}zquez} \& {Leitherer}(2005)}]{VazLei05}
{V{\'a}zquez}, G.~A. \& {Leitherer}, C. 2005, \apj, 621, 695

\bibitem[{{Veilleux} {et~al.}(2005){Veilleux}, {Cecil}, \&
  {Bland-Hawthorn}}]{Vei+05}
{Veilleux}, S., {Cecil}, G., \& {Bland-Hawthorn}, J. 2005, \araa, 43, 769

\bibitem[{{Wisnioski} {et~al.}(2012){Wisnioski}, {Glazebrook}, {Blake},
  {Poole}, {Green}, {Wyder}, \& {Martin}}]{Wis+12}
{Wisnioski}, E., {Glazebrook}, K., {Blake}, C., {Poole}, G.~B., {Green}, A.~W.,
  {Wyder}, T., \& {Martin}, C. 2012, \mnras, 422, 3339

\bibitem[{{Wright} {et~al.}(2010){Wright}, {Larkin}, {Graham}, \&
  {Ma}}]{Wri+10}
{Wright}, S.~A., {Larkin}, J.~E., {Graham}, J.~R., \& {Ma}, C.-P. 2010, \apj,
  711, 1291

\bibitem[{{Wright} {et~al.}(2009){Wright}, {Larkin}, {Law}, {Steidel},
  {Shapley}, \& {Erb}}]{Wri+09}
{Wright}, S.~A., {Larkin}, J.~E., {Law}, D.~R., {Steidel}, C.~C., {Shapley},
  A.~E., \& {Erb}, D.~K. 2009, \apj, 699, 421

\bibitem[{{Wright} {et~al.}(2007)}]{Wri+07}
{Wright}, S.~A. {et~al.} 2007, \apj, 658, 78

\bibitem[{{Wuyts} {et~al.}(2012){Wuyts}, {F{\"o}rster Schreiber}, {Genzel},
  {Guo}, {Barro}, {Bell}, {Dekel}, {Faber}, {Ferguson}, {Giavalisco}, {Grogin},
  {Hathi}, {Huang}, {Kocevski}, {Koekemoer}, {Koo}, {Lotz}, {Lutz}, {McGrath},
  {Newman}, {Rosario}, {Saintonge}, {Tacconi}, {Weiner}, \& {van der
  Wel}}]{Wuy+12}
{Wuyts}, S., {F{\"o}rster Schreiber}, N.~M., {Genzel}, R., {Guo}, Y., {Barro},
  G., {Bell}, E.~F., {Dekel}, A., {Faber}, S.~M., {Ferguson}, H.~C.,
  {Giavalisco}, M., {Grogin}, N.~A., {Hathi}, N.~P., {Huang}, K.-H.,
  {Kocevski}, D.~D., {Koekemoer}, A.~M., {Koo}, D.~C., {Lotz}, J., {Lutz}, D.,
  {McGrath}, E., {Newman}, J.~A., {Rosario}, D., {Saintonge}, A., {Tacconi},
  L.~J., {Weiner}, B.~J., \& {van der Wel}, A. 2012, \apj, 753, 114

\bibitem[{{Wuyts} {et~al.}(2011){Wuyts}, {Forster Schreiber}, {van der Wel},
  {Magnelli}, {Guo}, {Genzel}, {Lutz}, {Aussel}, {Berta}, {Cava},
  {Gracia-Carpio}, {Kocevski}, {Koekemoer}, {Lee}, {Le Floc'h}, {McGrath},
  {Nordon}, {Popesso}, {Pozzi}, {Riguccini}, {Rodighiero}, {Saintonge}, \&
  {Tacconi}}]{Wuy+11b}
{Wuyts}, S., {Forster Schreiber}, N.~M., {van der Wel}, A., {Magnelli}, B.,
  {Guo}, Y., {Genzel}, R., {Lutz}, D., {Aussel}, H., {Berta}, S., {Cava}, A.,
  {Gracia-Carpio}, J., {Kocevski}, D.~D., {Koekemoer}, A.~M., {Lee}, K.-S., {Le
  Floc'h}, E., {McGrath}, E.~J., {Nordon}, R., {Popesso}, P., {Pozzi}, F.,
  {Riguccini}, L., {Rodighiero}, G., {Saintonge}, A., \& {Tacconi}, L. 2011,
  ArXiv e-prints

\bibitem[{{Wuyts} {et~al.}(2008){Wuyts}, {Labb{\'e}}, {Schreiber}, {Franx},
  {Rudnick}, {Brammer}, \& {van Dokkum}}]{Wuy+08}
{Wuyts}, S., {Labb{\'e}}, I., {Schreiber}, N.~M.~F., {Franx}, M., {Rudnick},
  G., {Brammer}, G.~B., \& {van Dokkum}, P.~G. 2008, \apj, 682, 985

\bibitem[{{Xue} {et~al.}(2011){Xue}, {Luo}, {Brandt}, {Bauer}, {Lehmer},
  {Broos}, {Schneider}, {Alexander}, {Brusa}, {Comastri}, {Fabian}, {Gilli},
  {Hasinger}, {Hornschemeier}, {Koekemoer}, {Liu}, {Mainieri}, {Paolillo},
  {Rafferty}, {Rosati}, {Shemmer}, {Silverman}, {Smail}, {Tozzi}, \&
  {Vignali}}]{Xue+11}
{Xue}, Y.~Q., {Luo}, B., {Brandt}, W.~N., {Bauer}, F.~E., {Lehmer}, B.~D.,
  {Broos}, P.~S., {Schneider}, D.~P., {Alexander}, D.~M., {Brusa}, M.,
  {Comastri}, A., {Fabian}, A.~C., {Gilli}, R., {Hasinger}, G.,
  {Hornschemeier}, A.~E., {Koekemoer}, A., {Liu}, T., {Mainieri}, V.,
  {Paolillo}, M., {Rafferty}, D.~A., {Rosati}, P., {Shemmer}, O., {Silverman},
  J.~D., {Smail}, I., {Tozzi}, P., \& {Vignali}, C. 2011, \apjs, 195, 10

\bibitem[{{Yabe} {et~al.}(2012){Yabe}, {Ohta}, {Iwamuro}, {Yuma}, {Akiyama},
  {Tamura}, {Kimura}, {Takato}, {Moritani}, {Sumiyoshi}, {Maihara},
  {Silverman}, {Dalton}, {Lewis}, {Bonfield}, {Lee}, {Curtis Lake}, {Macaulay},
  \& {Clarke}}]{Yab+12}
{Yabe}, K., {Ohta}, K., {Iwamuro}, F., {Yuma}, S., {Akiyama}, M., {Tamura}, N.,
  {Kimura}, M., {Takato}, N., {Moritani}, Y., {Sumiyoshi}, M., {Maihara}, T.,
  {Silverman}, J., {Dalton}, G., {Lewis}, I., {Bonfield}, D., {Lee}, H.,
  {Curtis Lake}, E., {Macaulay}, E., \& {Clarke}, F. 2012, \pasj, 64, 60

\bibitem[{{Yuan} {et~al.}(2013){Yuan}, {Kewley}, \& {Richard}}]{Yua+13}
{Yuan}, T.-T., {Kewley}, L.~J., \& {Richard}, J. 2013, \apj, 763, 9

\end{thebibliography}

\nocite{PetPag04}
\nocite{KewEll08}
\nocite{Liu+08}
\nocite{For+13a}
\nocite{For+13b}
\nocite{Kur+13}
\nocite{Kau+03}
\nocite{Tre+04}
\nocite{Erb+06}
\nocite{Erb+06a}
\nocite{Jun+11}
\nocite{Jun+13}
\nocite{Tru+13}
\nocite{Wri+09}
\nocite{Wri+10}
\nocite{Shapley+05}
\nocite{New+12a}
\nocite{Gen+08}
\nocite{Gen+11}
\nocite{Gen+06}
\nocite{Duf+80}
\nocite{Dav+11}
\nocite{Wuy+12}
\nocite{BruCha03}
\nocite{Pen+02}
\nocite{For+09}
\nocite{For+11a}
\nocite{Man+11}
\nocite{vDok+05}
\nocite{Dad+04}
\nocite{Ste+04}
\nocite{Ade+04}
\nocite{Law+09}
\nocite{Abr+04}
\nocite{Mig+05}
\nocite{Kon+06}
\nocite{Kom+11}
\nocite{Cha03}
\nocite{Bri+08}
\nocite{BPT81}
\nocite{CowHuSon95}
\nocite{vdBer+96}
\nocite{ElmElmShe04}
\nocite{Elm+09}
\nocite{ElmElm05}
\nocite{ElmElm06}
\nocite{Law+07}
\nocite{Law+12}
\nocite{Swi+12a}
\nocite{Swi+12b}
\nocite{For+06}
\nocite{Wri+07}
\nocite{Sha+08}
\nocite{Cre+09}
\nocite{vSta+08}
\nocite{Epi+09}
\nocite{LemLam10}
\nocite{Jon+12}
\nocite{Wis+12}
\nocite{Swi+12a}
\nocite{Swi+12b}
\nocite{Epi+12}
\nocite{New+13a}
\nocite{Kri+07}
\nocite{Hay+09}
\nocite{Yab+12}
\nocite{Dav+07}
\nocite{For+11b}
\nocite{Kro01}
\nocite{Fer+13}
\nocite{HilMil98}
\nocite{Mey+94}
\nocite{Pau+01}
\nocite{Lei+99}
\nocite{VazLei05}
\nocite{GnaSte09}
\nocite{Asp+09}

\nocite{Kon+06}
\nocite{Ren+97}
\nocite{Stern+05}
\nocite{Cim+02}
\nocite{Xue+11}
\nocite{Mil+08}
\nocite{Bon+12} 
\nocite{Don+12}
\nocite{Lac+07}
\nocite{Kri+09}
\nocite{Bar+08}
\nocite{Ber+11}
\nocite{Sei+10}
\nocite{Age+10}
\nocite{Lil+07}
\nocite{Wuy+08}
\nocite{Ran+13}
\nocite{Bus+10}
\nocite{Yua+13}
\nocite{DopSut95}
\nocite{Dop+00}
\nocite{Dop+06}
\nocite{All+08}
\nocite{Lev+10}
\nocite{Ric+11}

\end{document}